\documentclass[conference,compsoc]{IEEEtran}
\IEEEoverridecommandlockouts
\usepackage{cite}
\usepackage{amsmath,amssymb,amsfonts}
\usepackage{algorithmic}
\usepackage{graphicx}
\usepackage{textcomp}
\usepackage{pgfplots}
\usepackage{pgf-pie}
\usepackage[table]{xcolor}
\usepackage{tabularx}
\usepackage{threeparttable}
\newcolumntype{C}{>{\centering\arraybackslash}X}
\usepackage{hyperref}
\usepackage{url}
\usepgfplotslibrary{statistics}
\pgfplotsset{compat=1.17}

\definecolor{lightgreen}{rgb}{0.85,1,0.85}
\definecolor{medgreen}{rgb}{0.6,0.9,0.6}
\definecolor{darkgreen}{rgb}{0.2,0.8,0.2}
\definecolor{lightred}{rgb}{1,0.8,0.8}
\definecolor{lightgray}{gray}{0.9}

\def\BibTeX{{\rm B\kern-.05em{\sc i\kern-.025em b}\kern-.08em
    T\kern-.1667em\lower.7ex\hbox{E}\kern-.125emX}}
\begin{document}

\title{X-amine509: Predicting the Practical Risk Level of Enterprise X.509 Certificates

% \thanks{Keyfactor}
}

% \author{\IEEEauthorblockN{Anonymous Authors}}

\author{\IEEEauthorblockN{Cameron Keith}
\IEEEauthorblockA{\textit{Analytics and AI Team} \\
\textit{Keyfactor}\\
Hanover, NH \\
cameron.s.keith.gr@dartmouth.edu}

\and
\IEEEauthorblockN{Shubh Patel}
\IEEEauthorblockA{\textit{Analytics and AI Team} \\
\textit{Keyfactor}\\
Cleveland, OH \\
shubh.patel@keyfactor.com}

\and
\IEEEauthorblockN{JD Kilgallin\textsuperscript{*}\thanks{\textsuperscript{*}Corresponding author: jd.kilgallin@keyfactor.com}}
\IEEEauthorblockA{\textit{Analytics and AI Team} \\
\textit{Keyfactor}\\
Cleveland, OH \\
jd.kilgallin@keyfactor.com}

\and
\IEEEauthorblockN{Caleb Shorter}
\IEEEauthorblockA{\textit{Analytics and AI Team} \\
\textit{Keyfactor}\\
Cleveland, OH \\
caleb.shorter@keyfactor.com}
}

\maketitle

\begin{abstract}
Enterprises managing large X.509 certificate inventories face a prioritization problem: deterministic analysis tools that precisely identify standards violations are indispensable for remediation, but applying them exhaustively across millions of certificates is operationally impractical. We present X-amine509, a two-stage triage system that uses machine learning to rapidly rank certificates by predicted risk and route only the highest-risk items to full deterministic analysis. Certificate risk is quantified as a composite score derived from 177 defect checks grounded in CA/Browser Forum Baseline Requirements, NIST IR~8547/SP~800-57, and cryptographic strength criteria, weighted by security severity across four tiers ranging from cryptographic breaks to minor compliance deviations. We collected 1{,}027{,}714 X.509 certificates from Fortune~500, .gov, and .edu domains and scored each using this rubric. On a held-out test set of 201{,}976 certificates, our best model (Extra Trees) achieves {\boldmath $R^2$} of 0.993 with MAE of 2.26, while Decision Tree scores {\boldmath $R^2$} of 0.986 at 3.7 million certificates per second on a single machine. Ranking quality confirms the triage value: aggregate NDCG exceeds 0.997, and severity-tier classification reports 99.76\% accuracy with 98.90\% recall on critical-tier defects. Thirteen months later, we retrieved another 571{,}374 certificates to test our models' durability over time, and the Extra Trees and Decision Tree models maintain MAE below 6.8, {\boldmath $R^2$} of at least 0.915, aggregate NDCG above 0.988, severity-tier accuracy of at least 99.52\%, and critical-tier recall of at least 97.03\%. Feature importance analysis identifies validity period, Extended Key Usage configuration, negative serial number encoding, and self-signed status as the strongest risk predictors, providing coarse interpretability at the triage stage.
\end{abstract}


\section{Introduction}
X.509 certificates are foundational to Internet security, because they enable authenticated key exchange in protocols such as TLS, which in turn protects the confidentiality and integrity of communications. An X.509 public key certificate is ``a set of data that uniquely identifies a public key (which has a corresponding private key) and an owner 
that is authorized to use the key pair. The certificate contains the owner's public key and possibly other information and is digitally signed
by a Certification Authority (i.e., a trusted party), thereby binding the public key to the owner'' \cite{raimondo2023fips}. Certificates play a vital role, but
misconfigured or weak certificates can expose enterprises to serious cybersecurity risks. High-profile incidents stemming from improperly configured or compromised certificates underscore the urgent need for effective risk assessment. For example, between 2015 and 2017, Symantec and its Registration Authority (RA) partners were caught issuing SSL/TLS certificates without proper domain ownership validation, including unauthorized Extended Validation certificates for Google-owned domains. Google's Certificate Transparency (CT) logs revealed that up to 30{,}000 certificates had been improperly issued across four RA partners that Symantec had failed to adequately audit. In response, Google Chrome and Mozilla Firefox executed a phased distrust of all Symantec-issued certificates by October 2018, which forced DigiCert (which acquired Symantec's CA business) to replace millions of certificates across the web \cite{kumar2018tracking}. In another example, the Equifax breach in 2017 
was exacerbated by an expired SSL certificate \cite{portman2019equifax}. 
In addition, malicious AI tools and probes are becoming more sophisticated and automated, and are able to identify and exploit many of these weaknesses \cite{paloalto2025ai}.

We identified four main categories of risks associated with X.509 certificates:
\begin{itemize}
    \item \textbf{Public Key Infrastructure (PKI) Chain Issues:} Chain validation failures, such as missing intermediate certificates, path length constraint violations, untrusted or unknown root CAs, and revoked certificates within the chain, undermine the CA trust architecture. These defects correspond to specific checks in x509lint \cite{roeckx2021x509lint} and RFC~5280 \cite{cooper2008rfc5280} chain validation requirements. When a chain certificate is compromised, attackers can access protected systems without authorization, inject malware, and collect sensitive consumer and enterprise data, leading to severe financial losses, government investigations, and heavy fines.
    \item \textbf{Key Compromise:} A certificate with a weak key is a critical vulnerability, because attackers can run a MITM attack, harvest credentials, or spoof trusted services.
    \item \textbf{Invalid Content:} The certificate structure or content is unusable for the intended purpose.
    \item \textbf{Visible Negligence:} These certificates fail to meet established best practices and call into question the trustworthiness of the PKI that issued them.
\end{itemize}

PKI administrators protect a hierarchy of cryptographic assets with varying criticality. The most critical is the certificate authority's (CA) private key, because compromise of this key allows an attacker to issue fraudulent certificates and undermine the entire trust hierarchy. Code-signing keys carry the next-highest impact, since a compromised code-signing key lets an attacker distribute signed malware that appears legitimate. Individual website certificate keys are lower-impact per certificate but far more numerous, making them a significant aggregate concern.

Certificates issued for phishing or spoofing domains represent a distinct category. These certificates are not issued by the enterprise, do not appear in the organization's certificate inventory, and require different response actions such as customer notification and domain takedown rather than certificate revocation. While phishing certificates may pose risks to an organization's customers, they do not reflect the security posture of the enterprise's own PKI operations.

Despite this distinction, most academic research on X.509 certificate risk focuses on phishing and malware classification, leaving practitioners without ML-assisted tools for prioritizing remediation across large enterprise certificate inventories. The methodological insights from phishing-focused work nonetheless inform our approach to building an enterprise risk model.

Traditional approaches to evaluate certificate risk depend on deterministic checks: static analyses of certificate attributes such as
expiry dates, cryptographic algorithms, and adherence to standards. Online tools like Bad Keys \cite{badkeys2025} and CRT \cite{sectigo2025} provide user interfaces to check individual certificates for issues.
While these deterministic checks are reliable for any individual certificate, at enterprise scale, the economics shift substantially.
The Keyfactor PKI \& Digital Trust Report \cite{keyfactor2024pki} reports that the average large enterprise manages 81{,}139 internally trusted certificates,
with the largest organizations managing hundreds of thousands or more.
In our experiments, deterministic analysis of 1{,}027{,}714 certificates required over 24 hours on comparable hardware, corresponding to approximately 0.1 seconds per certificate.
At this rate, a 10-million-certificate enterprise would require approximately ten days of continuous single-threaded analysis.
The CA/Browser Forum's approved transition to 47-day maximum certificate validity \cite{cabforum2025sc081} will cause certificate populations to turn over faster,
compounding this assessment burden.

These timescales conflict with operational requirements: quarterly compliance reviews must leave time for remediation after assessment,
and incident response (such as identifying certificates affected by a newly disclosed CA compromise) demands inventory-wide triage in hours, not weeks.
The question is not whether deterministic analysis is feasible given sufficient time and hardware; it is whether organizations should expend compute-days
exhaustively analyzing a certificate population that is overwhelmingly compliant when ML triage can rank the entire inventory in seconds
and direct deterministic resources to the certificates most likely to warrant investigation.
In our dataset, over 80\% of certificates had zero identified risk factors; investing deterministic cycles to confirm this absence is the inefficiency that triage addresses.

We address this gap by introducing a two-stage triage framework: a fast ML model ranks certificates by predicted risk, and deterministic tools then provide detailed, explainable assessment starting from the top of the ranked list. The ML model does not replace deterministic analysis; it prioritizes the order in which certificates receive it. All actionable reports, including specific rule violations and remediation guidance, come from the deterministic checker. Practitioners receive the same detailed explanations they expect; they simply receive the most critical reports first. A fast, well-aligned model turns easily observed certificate attributes into a proxy for a full suite of deterministic assessments, providing a practical path to manage compute costs while surfacing risk and raising the baseline of PKI hygiene in large organizations.

Our approach is grounded in industry-standard PKI requirements and security best practices. We evaluate certificates against defects defined by 
established standards such as RFC 5280 \cite{cooper2008rfc5280} and CA/Browser Forum requirements \cite{cabforum2024baseline}, assigning risk weights based on perceived security severity. 
A key focus of our research is systematically measuring how different preprocessing techniques, feature engineering strategies, and 
modeling approaches affect prediction performance, providing insights into which improvements offer the greatest value for certificate risk assessment.
\newline \indent Our main contributions are as follows:
\begin{enumerate}
    \item We collected over 1 million publicly accessible X.509 certificates from high-impact commercial and government domains, and we analyzed each against 177 deterministic defect checks with both custom-built and open-source tools to classify the risk level of certificates and determine the prevalence of a wide range of certificate risk factors.
    \item We explored a variety of problem formulations and approaches to devise a technique for classifying certificate risk without relying on the resource-intensive, exhaustive checks for individual risk factors. Critically, we systematically measured the effect of different 
      preprocessing, feature engineering, and modeling improvements to understand which techniques contribute most to better model performance. Our risk scoring is 
      grounded in defects defined by established PKI standards and best practices, with weights assigned based on security severity.

    \item We optimized a series of ML models to rapidly and accurately predict the risk level of any given certificate. Our optimal model in terms of the performance-speed tradeoff processes 3.7 million certificates per second on a single machine while achieving an {\boldmath $R^2$} of 0.986 on a test set of 201{,}976 certificates, with aggregate NDCG above 0.997, severity-tier classification of 99.76\% and 98.90\% recall on critical-tier defects. For an enterprise where fewer than 20\% of certificates have any identified risk factor, ML triage avoids spending over 80\% of deterministic analysis cycles confirming the absence of issues.
    \item  We scored 571{,}374 certificates collected thirteen months later without model retraining to test our model duraibility. Extra Trees and Decision Tree retain MAE below 6.8, {\boldmath $R^2$} of at least 0.915, aggregate NDCG above 0.988, severity-tier accuracy of at least 99.52\%, and critical-tier recall of at least 97.03\%.

\end{enumerate}

This paper proceeds as follows: Section 2 reviews related literature and existing research gaps. Section 3 details our ML pipeline methodology. 
Section 4 describes our experimental setup, results, and analysis. Section 5 discusses practical implications and recommendations for future research.

\section{Related Work}

This section reviews prior approaches to certificate risk assessment using machine learning. As established in Section~1, most prior work targets phishing and malware detection rather than enterprise PKI hygiene; nevertheless, these approaches inform our feature selection and modeling strategy. We organize the review into two categories: static field models, which rely on attributes encoded in the certificate itself, and context-aware models, which incorporate external signals such as domain reputation, certificate transparency logs, and issuer behavior.

\subsection{Static field models}

% Reference \cite{b5} wanted an automated way to predict how risky an X.509 certificate is in browser applications. For each certificate, \cite{b5} computed 43 features and used Random Forest to classify
% certificates as High, Medium, or Low risk. Reference \cite{b5} trained their model on 2,000 trusted certificates from Alexa domains and 900 untrusted certificates from phishtank.com.
% They utilize a rule-based approach to label their data into three categories, and trained the Random Forest model to predict those categories. However, they do not 
% provide the accuracy of their model, and they train their model on a small dataset. 

Static field models treat the certificate as the primary evidence source: key parameters, validity, extensions, subject/issuer fields, and chain properties.
Prior work shows that structure and syntax carry signal for security posture (e.g., weak keys, missing or incorrect extensions, unusual subject fields).
However, as noted in Section~1, this line of work primarily targets phishing detection rather than enterprise operational hygiene assessment.
For example, Reference \cite{dong2015beyond} developed a real-time phishing detection framework using certificate features. 
They trained a set of binary classifiers on 42 features to predict phishing and non-phishing websites.
They extracted multiple features from certificates (e.g., issuer, subject, validity period) and trained various ML models including Random Forest, Naive Bayes, Logistic Regression, and 
k-nearest neighbors to identify phishing websites using only the certificate and TLS information. Random Forest attained 95\% accuracy.
Certificate Authorities (CAs) enforce domain validation and vetting checks to block fraudulent issuances, but such measures offer little insight into the configuration 
and security posture of an organization's existing certificates.
While knowledge of phishing domains can help an organization protect its customers, it does not help
corporate PKI administrators assess the health of their own PKI operations or enterprise network security posture.
In addition, a binary classifier fails to prioritize the riskiest certificates for remediation.

In another example, Reference \cite{dong2016detection} approached rogue certificate detection by training deep neural networks to classify each certificate as ``rogue'' or ``benign'' based on features from X.509 fields.
Rogue certificates are valid, signed certificates issued to the wrong party (e.g., faulty domain validation, incorrect subject/SAN, or issuance in violation of Baseline Requirements) or
certificates whose private keys are compromised. 
This definition does not comprehensively capture the impact on enterprise systems of privately trusted certificates with keys that are not known to be compromised.
Additionally, certificates with compromised keys are not a rigorously defined set, and this technique does not assist an administrator with determining if a key has been compromised.
The authors trained one model per root CA on 889{,}849 benign certificates from Alexa's top one million sites \cite{alexa2020top1m} plus 2{,}516{,}155 user-submitted certificates,
augmented with a few hundred real and synthetic rogue examples. Their classifiers reached 99\% accuracy on the top ten CAs. 
However, similar to \cite{dong2015beyond}, treating risk as a binary classification ignores the spectrum of threat levels and offers no continuous risk score for prioritization.

Reference \cite{li2021machine} built a Verification for Extraction (VFE) system to pull 182 certificate attributes and then tested classical (LR, DT, SVM), 
ensemble (Random Forest, XGBoost, LightGBM, CatBoost), and deep learning (e.g., CNN, LSTM) models to classify certificates as malicious or benign. The VFE system is a process 
that analyzes certificates' basic fields, checks whether each certificate conforms to constraints specified in RFC 5280 \cite{cooper2008rfc5280} (which is a memo on Internet X.509 PKI certificate and Certificate Revocation List (CRL) Profile), 
and constructs and verifies the certificate chain.
They labeled certificates as malicious if they were found in phishing or malware datasets. 
They trained on 11{,}610 samples from PhishTank URLs \cite{phishtank2025}, the SSL Blacklist project \cite{sslblacklist2025}, and the Alexa top-1 million sites \cite{alexa2020top1m}. 
Nonetheless, PhishTank \cite{phishtank2025} and SSL Blacklist certificates \cite{sslblacklist2025} offer only a partial view of enterprise risk, because they only aim to identify certificates related to phishing activity and botnet C\&C 
services and do not provide insights on privately trusted certificates used in enterprise environments. 
Using those 182 numerical features, their SVM detector hit 98.2\% accuracy and the ensemble models averaged 95.9\%.
While they achieved higher accuracy on their dataset, it is unclear whether the models will generalize to new certificates, because they were trained on a small dataset.
Similar to \cite{dong2015beyond, dong2016detection}, phishing and malware certificates do not provide a comprehensive way to evaluate enterprise PKI risk. 

Reference \cite{liu2022malcertificate} used a graph convolutional network to classify certificates as malicious or benign. Instead of manual feature engineering, they model each certificate as a heterogeneous graph: 
nodes represent attributes and edges represent relationships in the same certificate. They trained this model on 8061 certificates 
(6438 benign from Alexa's top 1 million domains \cite{alexa2020top1m} and 1623 malicious from abuse.ch \cite{sslblacklist2025}) after OpenSSL decryption and cleaning.
Abuse.ch is a community platform that tracks malware and botnets; its SSL Blacklist \cite{sslblacklist2025} lists TLS certificates and JA3 fingerprints observed in C2 infrastructure.
Cert GCN obtained 97.41\% accuracy on malicious samples and 92.89\% accuracy on benign samples. 
However, this non-traditional representation of certificates on a small dataset may not generalize well to all certificate types, use cases, and new potential defects. 

\subsection{Context-aware models}

Context-aware models incorporate signals beyond the certificate itself, such as how and where the certificate is used, properties of the domain or server presenting it, and time-based or network-wide patterns.
However, much of this work is misaligned with enterprise PKI operations, because its scope is limited to classifying certificates as phishing or malware certificates.

\textbf{Domain Name Features: } A TLS server-authentication certificate's risk is often tied to the domain name for which it was issued.
Reference \cite{hageman2022detecting} proposes a two-stage pipeline that extracts 107 features from X.509 fields and SAN lexical properties.
They trained multi-class classifiers and regression models (using algorithms such as \textit{Random Forest}) to assign both discrete risk categories (Phishing, Benign, and Conflicted)
and continuous ``phishiness'' scores from zero to one.
They collected 275{,}187 certificates from Tranco top-million list and eCrime Exchange (eCX) phishing URLs through Censys.
An eCX phishing URL is a phishing link reported within the Anti-Phishing Working Group platform, which consists of vetted members, and is shared via eCX for defenders to use.
They also performed a time-series analysis and showed that their model generalizes over time on their training dataset.
However, a credible enterprise certificate has no phishing domains, and a phishing domain appearing on an enterprise certificate should be considered a security incident. 
Also, the number of phishing domains does not necessarily correlate with organizations' overall risk due to phishing attacks.
% Additionally, they assign regression labels as the simple ratio of phishing SANs to total SANs for each certificate, which treats every domain with equal weight and 
% dilutes the practical risk signal when many SANs lack labels. This approach underplays cases where a single high-impact phishing domain should dominate the score, so you can't trust 
% the ``phishyness'' values as meaningful risk estimates.

\textbf{Certificate Transparency (CT) Log Mining: } Reference \cite{fasllija2019phish} built a three-stage pipeline (Certificate Collector, Feature Extractor, and Classifier) 
that extracts eight lightweight features (e.g., domain similarity, nested subdomains, free CA indicators, suspicious top-level domains (TLDs), inner TLD in the subdomain, high Shannon entropy, hyphens in the subdomain) 
and trained ML models to classify certificates as ``legitimate'', ``potential'', ``likely'', ``suspicious'', and ``highly suspicious'' (SVM, Decision Tree, k-NN, MLP).
Domain similarity is based on the edit distance of a website domain name to a small corpus of suspicious keywords. High Shannon entropy is a measure of the degree of randomness in a domain name.
The authors argue that a domain with high entropy may indicate that it is a phishing domain. The authors trained on a dataset of roughly 
600{,}000 certificates collected via the CertStream library from public CT logs and labeled by a heuristic scoring method. They used eight features to train their models. 
Their SVM and Decision Tree classifiers recorded over 90\% accuracy and F1 scores in near-real-time detection. The minimal feature set may omit important signals of risk such as
validity period and extensions present. Therefore, the models will fail to detect issues not captured by these features. If an attacker wanted to exploit this ML framework, 
they would simply have to modify a couple of elements of the certificate to bypass a suspicious classification. For example, an attacker could:
\begin{itemize}
  \item Use a paid CA.
  \item Avoid the following: 
    \begin{itemize}
      \item Suspicious TLDs
      \item Inner TLDs in the subdomain
      \item Deeply nested subdomains
      \item Multiple hyphens in the domain
    \end{itemize}
\end{itemize}

% \textbf{Domain Reputation and Popularity:} Reference \cite{li2021machine} also segmented benign certificates by the Alexa rank of their domains to analyze performance. 

\textbf{Issuer Behavior and CA Reputation:} Reference \cite{hageman2021can} implemented a Certificate Transparency log analysis pipeline that samples 30{,}000 domains (phishing, popular benign, non-popular benign), 
pulls all matching certificates via Censys, 
and computes statistical metrics on issuer choice, validation level, and certificate sharing. They analyzed a dataset of 79.1 million certificates 
collected in February-March 2020 from CT logs via Google BigQuery. Rather than a classic feature vector, they derived three main feature groups: issuer preferences across 853 authorities, 
validation level categories (Domain/Organization/Extended Validation (DV/OV/EV), or unknown), and counts of unique root domains per certificate. 
Their results show that 56-75\% of phishing domains have certificates issued before 
blacklisting, phishers concentrate on a small set of CAs (notably cPanel), and EV certificates are almost unused. 
While they have a large dataset and conduct analysis on it,
they did not train any models to predict risk.

\textbf{Emerging Techniques with LLMs:} Reference \cite{rapid72024llm} leverages pre-trained LLMs to circumvent traditional feature engineering. Instead of 
training a classifier from scratch, they harness LLMs to embed the certificate's subject and issuer names into a semantic vector space. 
Then, they utilize k-nearest neighbors to find the most similar known certificates and vote on whether the new certificate is malicious. 
They scored 99.4\% accuracy with the OpenAI embedding model. This approach may reduce the need for maintaining a custom-trained model, because it relies on natural language to classify certificates.
However, relying on OpenAI embeddings adds latency, recurring costs, and an opaque black box, and LLMs are much more expensive than an end-to-end ML pipeline to predict risk.
Furthermore, they lack an ablation study to analyze which parts of
the certificate subject and issuer names contribute to higher performance. 

While both strategies have identified useful signals to detect malicious, rogue, or phishing domains, 
these approaches are not designed to predict risk in enterprise PKI environments. 
References \cite{dong2015beyond,li2021machine,liu2022malcertificate} rely on limited datasets.
Nearly all reduce risk to a binary or coarse three-level label, which masks medium-risk cases and prevents one from ranking threats. 
While \cite{hageman2022detecting} trained a regression model, phishing domains do not appear in enterprise certificates, so this kind of ML model cannot predict enterprise risk well.
More complex approaches such as deep neural networks in \cite{dong2016detection,li2021machine}, graph models in \cite{liu2022malcertificate}, and embedding-based classification \cite{rapid72024llm}
add training overhead without offering significant improvements in performance, interpretability, or generalization, nor providing a continuous risk score.

Table~\ref{tab:comparison} summarizes how our work differs from prior certificate analysis research. These are fundamentally different problems that happen to share some overlapping features (key parameters, extensions, subject fields); the differences in problem scope, output type, ground truth, and scale each motivate distinct modeling choices.

\begin{table}[h]
\caption{Comparison with Prior Certificate Analysis Research}
\centering
\label{tab:comparison}
\setlength{\tabcolsep}{3pt}
\footnotesize
\begin{tabular}{|l|l|l|}
\hline
\textbf{Dimension} & \textbf{Prior Work} & \textbf{This Work} \\
\hline
\parbox[t]{1.2cm}{Problem\\scope} & \parbox[t]{3cm}{\raggedright Phishing/malware detection: is this cert associated with a malicious site?} & \parbox[t]{3cm}{\raggedright Enterprise hygiene: how well-managed is this cert relative to security policies?} \\
\hline
\parbox[t]{1.2cm}{Output\\type} & \parbox[t]{3cm}{\raggedright Binary or coarse (benign/malicious, 3-level) classification} & \parbox[t]{3cm}{\raggedright Continuous risk score enabling fine-grained ranking for triage} \\
\hline
\parbox[t]{1.2cm}{Ground\\truth} & \parbox[t]{3cm}{\raggedright Blacklist labels (e.g., PhishTank, VirusTotal)} & \parbox[t]{3cm}{\raggedright Standards-based rubric from RFC~5280, CA/B~BRs, NIST} \\
\hline
Scale & \parbox[t]{3cm}{\raggedright Corpora reach tens of millions, but labeled by external reputation signals or heuristics} & \parbox[t]{3cm}{\raggedright 1{,}027{,}714 real-world certs from Fortune~500, .gov, .edu, each scored by 177 deterministic defect checks} \\
\hline
\end{tabular}
\end{table}

This paper addresses that gap by modeling \textit{practical} enterprise PKI risk from certificate-visible attributes. Rather than
defining risk heuristically or based on phishing datasets, we ground our defect detection in established PKI standards such as RFC 5280 \cite{cooper2008rfc5280}
and CA/Browser Forum baseline requirements \cite{cabforum2024baseline}, assigning risk weights based on security severity.

Beyond the mismatch in problem scope, prior work also offers limited insight into which modeling choices matter most for performance. 
Few studies systematically ablate their design decisions to understand the relative contribution of feature engineering, model selection, 
or hyperparameter tuning. Our work addresses this by explicitly measuring how different preprocessing techniques, feature engineering 
strategies, and modeling approaches affect model performance, providing practitioners with actionable guidance on where to invest effort 
when building similar risk assessment systems.

Finally, most prior work does not address computational efficiency at scale. Enterprise environments with millions of certificates 
require not just accurate risk assessment, but assessment that can be performed rapidly enough to enable continuous monitoring and 
real-time decision-making. Our approach is several orders of magnitude faster than traditional deterministic analysis while maintaining 
high model performance.

\section{Methodology}\label{sec:method}
Our methodology addresses two objectives: predicting X.509 certificate risk efficiently and measuring which modeling choices contribute most to performance.

We define \textit{certificate risk} as a composite measure of the degree and severity of deviation from established PKI standards and security best practices, as assessed by a deterministic rubric derived from RFC~5280 \cite{cooper2008rfc5280}, CA/Browser Forum Baseline Requirements \cite{cabforum2024baseline}, and NIST guidelines \cite{nist2024ir8547, nist2020sp80057}. Higher scores indicate more numerous and/or more severe deviations. This operational definition captures \textit{indicators} of risk rather than a probabilistic risk estimate, as the probability and impact of exploitation depend on deployment context not captured in the certificate itself.

We evaluate certificates against 177 defect criteria derived from these standards,
spanning four risk categories: PKI chain issues, key compromise indicators, invalid content, and visible negligence. Each defect receives
a severity weight, and we compute a composite risk score as our prediction target.

Using open-source tools and implementations developed by the research team, we analyzed collected certificates for the following defects and scored 
each finding on a certificate based on a rough rubric of severity. The team iterated on several sets of scoring weights until results matched expectations 
from a panel of enterprise PKI practitioners and security solution vendor personnel. It is likely that further refinement of the scores would provide even 
more robust results, but our conclusions are similar under all weights tested: Certificates with scores over 1000 contain at least one critical risk, or so
many non-fatal but serious risks that they should not be relied on. Our ML models learn to efficiently approximate the output of this standards-based deterministic rubric. The contribution is the demonstrated feasibility of this approximation at scale with sufficient ranking quality to enable triage, not the rubric itself.
The rubric represents one operationalization of certificate risk; organizations may adapt it to their specific policies.

\begin{itemize}
  \item More than 100 checks implemented in the x509lint certificate linting program \cite{roeckx2021x509lint}, identifying failures of RFC 5280 and CA/B Forum Baseline Requirements.
  \begin{itemize}
  \item O(1000) points: Fatal issue -- key or CA fundamentally insecure
  \item O(100) points: Failure of a firm requirement on applicable certificate with no direct exploitability
  \item O(10) points: Failure of best practice
  \item O(1) points: Non-conventional but within normal operation/low risk
  \end{itemize}

  \item Certificates failing to conform to NIST IR 8547 \cite{nist2024ir8547} or NIST SP 800-57 \cite{nist2020sp80057} based on expiration date
  \begin{itemize}
  \item 1-100 points scaling based on bits of security and minimum key strength guidelines in effect at time of expiration
  \end{itemize}

  \item Outliers for validity period, SAN count, and file size based on Section~\ref{sec:eda} described below, which are known to cause incompatibilities with some systems
  \begin{itemize}
  \item 1000 points: Certificates $>$ 100~KB and known to be incompatible with many OpenSSL-based deployments \cite{openssl2015maxcert}. Certificates that are inconsistently recognized by monitoring tools represent intrinsic risk.
  \item 1-100 points: Validity/SAN/filesize outside normal range, scaling by magnitude of deviation
  \item 1000 points: RSA certificates with a modulus factorable using known techniques from prior work \cite{kilgallin2019factoring, heninger2012mining}
  \end{itemize}
\end{itemize}

To illustrate how the rubric maps defects to scores, we present three worked examples spanning different severity levels:

\textbf{Critical (O(1000)): Factorizable RSA modulus.}
An RSA certificate whose public modulus shares a prime factor with another certificate, recoverable via GCD-based techniques \cite{heninger2012mining}, receives 1000 points. This violates the fundamental assumption of RSA security (that factoring the modulus is computationally infeasible) and constitutes a complete cryptographic break: an attacker who recovers the private key can impersonate the certificate holder, forge signatures, and decrypt intercepted traffic. The score is assigned at the highest tier because the defect is directly exploitable with no additional conditions.

\textbf{Significant (O(100)): Missing Extended Key Usage on an end-entity certificate.}
An end-entity certificate that omits the Extended Key Usage (EKU) extension receives 100 points. CA/Browser Forum Baseline Requirements (Section 7.1.2.3) \cite{cabforum2024baseline} mandate EKU on subscriber certificates. Without this constraint, a certificate issued for TLS server authentication could also be accepted for code signing, email signing, or client authentication by relying party software that does not independently enforce usage restrictions. The score is O(100) rather than O(1000) because the defect enables potential misuse but does not constitute a direct cryptographic compromise; exploitation depends on the deployment context and the behavior of relying party software.

\textbf{Minor (O(10)): Non-standard encoding in a non-critical field.}
A certificate with a bit string containing a leading zero in a non-critical extension receives 10 points. This violates the Distinguished Encoding Rules (DER) specified in RFC~5280 \cite{cooper2008rfc5280} Section~4.1, which require a unique encoding for each ASN.1 value. The defect is not directly exploitable but indicates that the issuing CA uses non-compliant software, which correlates with other hygiene failures (Table~\ref{tab4}). The score is O(10) because the deviation is a best-practice failure with no security consequence in isolation.

These traditional deterministic analyses check each certificate
against all criteria, which is a computationally expensive process. Certificate parsing from ASN.1 DER or PEM format typically completes in milliseconds; the computational expense arises from 177 checks performed after parsing, particularly cryptographic analyses such as RSA modulus factorization \cite{kilgallin2019factoring, heninger2012mining}, which dominates processing time. Embedding all deterministic checks during the parsing phase would not meaningfully reduce per-certificate cost because parsing is not the bottleneck. Our goal is to predict these risk scores from certificate-visible attributes alone,
enabling assessment at millions of certificates per second.

\subsection{Data Collection}
We retrieved X.509 certificates from Certificate Transparency (CT) logs and host scan records from Keyfactor Command Risk Intelligence, 
collecting up to 1000 certificates each from 61{,}042 domains. As presented in Fig.~\ref{fig1}, our dataset comprised certificates from 
Fortune 500 companies (80.7\%), .gov domains (17.6\%), and .edu domains (1.7\%). 
This comprehensive list includes all known domains associated with contemporary Fortune 500 companies and .gov or .edu top-level domains (TLDs), 
based on reverse-whois and domain registration search.
Fig.~\ref{fig2} shows the trust distribution: 
85.6\% were publicly trusted certificates, 8.2\% were self-signed, and 6.2\% were privately rooted certificates. As displayed in Table~\ref{tab1}, 
we collected 453{,}917 certificates on June 17--18, 2024, and an additional 50{,}819 certificates on July 24, 2024.
Within the span of June 20--30, 2025, we collected another 522{,}978 certificates from the same domains, bringing the modeling corpus total to 1{,}027{,}714 certificates. During August 5--9, 2026, we collected 571{,}374 more certificates from the same domains, which brought the total to 1{,}599{,}088 certificates. Also, we hold out the certificates from August 5--9, 2026 as a post-deployment set to test temporal model durability in Section~\ref{sec:postdeploy}; it contributes to no distribution or statistics detailed in this section, to the exploratory data analysis (EDA) in Section~\ref{sec:eda}, or to model training.

This trust distribution has implications for our rubric's applicability. CA/Browser Forum Baseline Requirements directly govern the 85.6\% of certificates that are publicly trusted; publicly trusted CAs must comply with these requirements as a condition of inclusion in browser and operating system root stores. For the remaining 14.4\% (self-signed and privately rooted certificates), CA/B BRs are not legally binding. However, many enterprises adopt BR-derived policies as internal baselines even for private PKI, because these requirements codify widely accepted security practices such as maximum validity periods, minimum key sizes, and required extensions. More fundamentally, RFC~5280 \cite{cooper2008rfc5280}, which underpins the majority of our checks, defines the X.509 certificate profile independent of trust model and applies to all certificates regardless of whether they are publicly or privately rooted. We acknowledge that private CAs may legitimately deviate from BR-specific requirements (e.g., longer validity periods for internal services or different EKU usage patterns), and we discuss this as a generalizability consideration in Section~5.

\begin{figure}[h]
  \centering
  \includegraphics[width=1\linewidth]{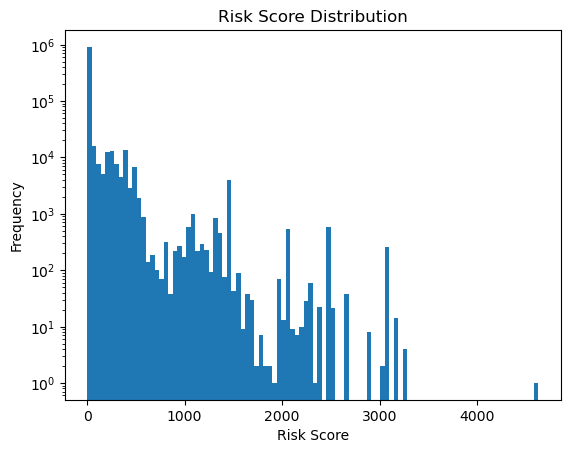}
  \caption{Practical Risk Score Distribution}
  \label{fig3}
\end{figure}

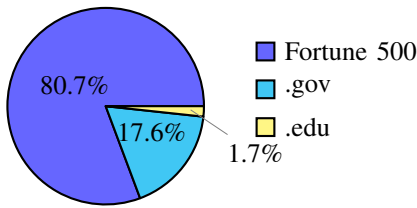
\begin{figure}[h]
  \centering
  \vspace{1mm}
  \begin{tikzpicture}
    \pie[
      text=legend,
      hide number,
      sum=100,
      radius=1.3,
      after number=\%,
      /tikz/every pin/.style={font=\footnotesize}
    ]{
      80.7/Fortune 500,
      17.6/.gov,
      1.7/.edu
    }
    \node at (145.26:0.5) {80.7\%};
    \node at (330.20:0.7) {17.6\%};
    \node[pin=-40:1.7\%]
      at (2:1) {};
  \end{tikzpicture}
  \caption{Domain Type Distribution}
  \label{fig1}
\end{figure}

\begin{figure}[h]
  \centering
  \vspace{1mm}
  \begin{tikzpicture}
    \pie[
      text=legend,
      hide number,
      sum=100,
      radius=1.3,
      after number=\%,
      /tikz/every pin/.style={font=\footnotesize}
    ]{
      85.6/Public,
      8.2/Self-signed,
      6.2/Private
    }
    \node at (160:0.5) {85.6\%};
    \node[pin=-40:8.2\%]
      at (-40:1) {};
    \node[pin=-20:6.2\%]
      at (-15:1) {};
  \end{tikzpicture}
  \caption{Certificate Type Distribution}
  \label{fig2}
\end{figure}
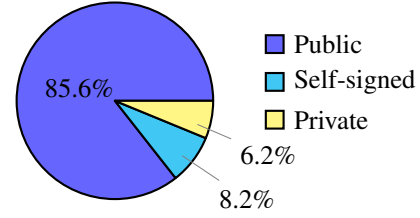

As shown in Fig.~\ref{fig3},
the risk scores range from 0 to 4627. The median risk score is 1, the mean is 41.31, and the standard deviation is 177.5. 

\begin{table}[h]
  \caption{Dates of Data Collection}
  \centering
  \begin{tabular}{|l|r|}
    \hline
    \textbf{Date}               & \textbf{Certificates} \\
    \hline
    June 17--18, 2024       & 453{,}917                    \\
     \hline
    July 24, 2024                     &  50{,}819                    \\
    \hline
    \textbf{Subtotal} & \textbf{504{,}736}           \\ 
    \hline
    June 20--30, 2025       &  522{,}978                   \\
    \hline
    \textbf{Modeling corpus} & \textbf{1{,}027{,}714}           \\
    \hline
    August 5--9, 2025       &  571{,}374                   \\
    \hline
    \textbf{Total collected} & \textbf{1{,}599{,}088}           \\
    \hline
  \end{tabular}
  \label{tab1}
\end{table}

\subsection{Data Preprocessing and Feature Extraction}
To prepare the data for modeling, we derived 69 features from the issuer, subject fields, usage and constraints, public key, and signature.
For example, some of these features are:

\begin{itemize}
  \item Key size in bits
  \item Key type
  \item Certificate version
  \item Number of SANs
  \item Validity period in months
  \item Is self-signed
  \item Subject fields: CN, SN, serial number, C, L, ST, street, O, OU, title, G, E, UID, DC, initials, generation qualifier, DN qualifier, pseudonym
  \item Other names: email address, DNS name, X.400 address, directory name, EDI party name, URI, IP address, registered ID, UPN, DS replication, unknown
  \item Key usage flags: no key usage, encipher only, CRL signing, key certificate signing, key agreement, data encipherment, key encipherment, non-repudiation, digital signature, decipher only
\end{itemize}

\subsection{Exploratory Data Analysis}\label{sec:eda}
After we processed the data, we performed an EDA to understand the distribution of features and identify potential correlations.
Expiration dates ranged from June 17, 2024, to December 31, 9999. The number of Subject Alternative Names (SANs) ranged from 0 to 4772.
File sizes in PEM encoding ranged from 1 KB to 87 KB. 508 public certificates did not meet CA/Browser Forum Baseline Requirements \cite{cabforum2024baseline}. 
91{,}239 certificates (18.1\% of the initial data collection) contained at least one identified risk factor. Table~\ref{tab2} illustrates the prevalence of select issues across our dataset.
For example, some of the most common certificate issues were no policy extension, no key usage, old certificate version, negative serial number, and RSA modulus smaller than 2048 bits.

Furthermore, as reflected in Table~\ref{tab4}, this correlation may stem from a common cause: insufficient validation of certificate requests by the issuing PKI prior to certificate issuance. This raises the natural question about whether such certificates can be characterized, and whether the risk level of an arbitrary certificate could
be easily predicted based on such a characterization.

For example, as shown in Table~\ref{tab2}, the most common are basic hygiene gaps: no policy extension (1/10), no key usage (1/25), 
non-v3 certs (1/26), weak RSA keys (1/35), and negative serial numbers (1/37). This pattern suggests poor profile enforcement at issuance,
with clear compliance and interoperability fallout. Mid-tier issues like duplicate SANs (1/151) show up occasionally, while formatting and 
metadata errors (invalid time formats (1/1550), leading-zero bit strings (1/1960), non-ISO country codes (1/2080), empty issuers (1/4950)) are relatively rare.

\begin{table}[htbp]
  \centering
  \begin{threeparttable}
    \caption{Correlation of Linting Defects}
    \label{tab4}

    % tighten padding
    \setlength{\tabcolsep}{1.5pt}
    \renewcommand{\arraystretch}{1.1}

    \begin{tabular}{|l|c|c|c|c|c|c|}
      \hline
      \shortstack[l]{No policy\\extension}
        & 1 & \cellcolor{lightgray} & \cellcolor{lightgray}
        & \cellcolor{lightgray} & \cellcolor{lightgray}
        & \cellcolor{lightgray} \\
      \hline
      Valid $>$ 60 months
        & \cellcolor{medgreen}0.81 & 1 & \cellcolor{lightgray}
        & \cellcolor{lightgray} & \cellcolor{lightgray}
        & \cellcolor{lightgray} \\
      \hline
      \shortstack[l]{No OCSP\\over HTTP}
        & \cellcolor{darkgreen}0.99 & \cellcolor{medgreen}0.81
        & 1 & \cellcolor{lightgray} & \cellcolor{lightgray}
        & \cellcolor{lightgray} \\
      \hline
      No AIA ext.
        & \cellcolor{darkgreen}0.96 & \cellcolor{medgreen}0.82
        & \cellcolor{darkgreen}0.97 & 1 & \cellcolor{lightgray}
        & \cellcolor{lightgray} \\
      \hline
      \shortstack[l]{No HTTP URL\\in AIA ext.\tnote{a}}
        & \cellcolor{darkgreen}0.98 & \cellcolor{medgreen}0.82
        & \cellcolor{darkgreen}0.98 & \cellcolor{darkgreen}0.98
        & 1 & \cellcolor{lightgray} \\
      \hline
      No EKU
        & \cellcolor{medgreen}0.74 & \cellcolor{medgreen}0.76
        & \cellcolor{medgreen}0.75 & \cellcolor{medgreen}0.75
        & \cellcolor{medgreen}0.75 & 1 \\
      \hline
      % final row: full vertical bars + rotated labels anchored at bottom right
      {} 
        & \rotatebox[origin=br]{75}{\strut No policy extension}
        & \rotatebox[origin=br]{75}{\strut Valid $>$60months}
        & \rotatebox[origin=br]{75}{\strut No OCSP over HTTP}
        & \rotatebox[origin=br]{75}{\strut No AIA ext.}
        & \rotatebox[origin=br]{75}{\strut No HTTP URL in AIA ext.\tnote{a}}
        & \rotatebox[origin=br]{75}{\strut No EKU} \\
      \hline
    \end{tabular}

    \begin{tablenotes}[flushleft]\footnotesize
      \item[a] The Authority Information Access (AIA) extension provides URLs for OCSP responders and CA certificate retrieval. Absence of an HTTP URL in this extension prevents automated certificate chain building and revocation checking.
    \end{tablenotes}
  \end{threeparttable}
\end{table}

\begin{table}[h]
  \caption{Certificate Issue Frequency}
  \centering
  \begin{tabular}{|l|r|}
    \hline
    \textbf{Certificate Issue}               & \textbf{1 in every:} \\
    \hline
    No Policy Extension       & 10                   \\
     \hline
    No Key Usage                     &  25                   \\
    \hline
    Certificate Not Version 3       &  26                  \\
    \hline
    RSA modulus smaller than 2048 bits       &  35                  \\
    \hline
    Negative Serial Number       &  37                  \\
    \hline
    Duplicate SAN Entry & 151 \\
    \hline
    Invalid Time Format & 1550 \\
    \hline
    Bit String with Leading Zero & 1960 \\
    \hline
    Country Name not 2 Chars Long & 2080 \\
    \hline
    Empty Issuer & 4950 \\
    \hline
    URL contains a null character & 13850 \\
    \hline
    Signature Algorithm Mismatch & 18890 \\
    \hline
  \end{tabular}
  \label{tab2}
\end{table}

While individual encoding defects such as negative serial numbers or leading-zero bit strings are not directly exploitable, their presence indicates that the issuing CA does not enforce standard-compliant certificate profiles. As Table~\ref{tab4} demonstrates, these defects are highly correlated ($r = 0.74$--$0.99$), suggesting that certificates exhibiting one hygiene failure are likely to exhibit multiple others. We therefore treat these defects as proxy signals for the overall quality of PKI management, not as individual vulnerabilities. This interpretation is consistent with enterprise security assessment practice, where compliance deviations serve as leading indicators of more serious underlying problems.

\section{Experiments}
After preprocessing and analyzing the data, we trained and evaluated a series of regression models to predict certificate risk scores. 
We systematically varied training techniques, model architectures, and hyperparameters to measure their individual and combined effects on performance. 
This section describes our experimental setup, including data splits, evaluation metrics, baseline models, and training procedures, 
followed by our results and analysis.

\subsection{Dataset Splits}
We processed the full dataset of 1{,}027{,}714 certificates, removing 17{,}835 certificates due to missing data or processing errors 
(e.g., certificates that could not be loaded from a raw PEM or DER encoding). As shown in Table~\ref{tab10}, we split the 
remaining 1{,}009{,}879 certificates into training (80\%, 807{,}903 certificates) and test (20\%, 201{,}976 certificates) sets.

\begin{table}[h]
  \caption{Dataset Splits}
  \centering
  \begin{tabular}{|l|r|}
    \hline
    \textbf{Data}               & \textbf{Certificates} \\
    \hline
    Raw dataset       & 1{,}027{,}714                    \\
     \hline
    Processed dataset &  1{,}009{,}879                    \\
    \hline
    Train split &  807{,}903                    \\
    \hline
    Test split &  201{,}976                    \\
    \hline
    Post-deployment split &  571{,}374                    \\
    \hline
  \end{tabular}
  \label{tab10}
\end{table}

\subsection{Evaluation Metrics}\label{sec:metrics}
We evaluated model performance using three standard regression metrics: Mean Absolute Error (MAE), Mean Squared Error (MSE),
and $R^2$ score. MAE measures the average absolute prediction error in risk score units. MSE penalizes larger errors more
heavily by squaring the differences. $R^2$ quantifies model fit, ranging from 0 (no explanatory power) to 1 (perfect fit).
We also measured end-to-end pipeline runtime per certificate in nanoseconds, including both preprocessing and prediction.

To evaluate ranking quality under the triage framing, we additionally report
normalized discounted cumulative gain (NDCG@$k$), which measures how well a
model preserves the ordering of the ground-truth risk scores and penalizes
misordered items near the top of the list more. Point-prediction metrics (MAE,
$R^2$) do not capture this, because a model can have low average error while
misordering the certificates that matter most.

Given a ranking of $N$ certificates produced by sorting predicted risk in
descending order, let $g_i$ denote the \emph{ground-truth} composite risk score
of the certificate placed at position $i$. Discounted Cumulative Gain at depth
$k$ is
\begin{equation}
\mathrm{DCG}@k = \sum_{i=1}^{k} \frac{g_i}{\log_2(i+1)},
\end{equation}
which credits each certificate by its true risk and discounts that credit
logarithmically with rank, so misplacing a high-risk certificate near the top of
the queue costs more than misplacing one further down the list. Normalizing by the
ideal ordering $\mathrm{IDCG}@k$, obtained by sorting the same certificates by
descending true score, gives
\begin{equation}
\mathrm{NDCG}@k = \frac{\mathrm{DCG}@k}{\mathrm{IDCG}@k},
\end{equation}
which equals 1 exactly when the top $k$ positions are ordered as the ground
truth requires.

We use linear gain, so relevance is the raw composite risk score rather than a
bucketed or exponentiated transform, and certificates with identical predicted
scores receive the average gain of their tied positions, which makes the metric
independent of tie-breaking order. We report NDCG at 720 logarithmically spaced
depths and refer to $\mathrm{NDCG}@N$, evaluated over the complete ranked list,
as the \emph{aggregate NDCG}. Aggregate NDCG is a single evaluation at $k = N$,
not an average of the NDCG@$k$ curve.

We also convert continuous predictions to severity tiers (Low, High Severity, Critical) and report tier classification accuracy, per-tier precision, recall, and F1 score.

\subsection{Baseline Models}
First, to gauge whether ML models could predict risk scores, we trained a set of scikit-learn models \cite{pedregosa2011scikit} on a smaller subset of 20{,}000 certificates (from the training set). 
As indicated in Table~\ref{tab9}, Decision Tree performed the best, and it achieved the lowest MAE and MSE, and the highest $R^2$ score among the baseline models with 19.92, 19907, and 0.563, respectively.
These baseline models scored surprisingly well on the test set with 100{,}000 certificates.

\begin{table}[h]
\centering
\begin{threeparttable}
\caption{Baseline 20K Model Results on the 100K Test Set}
\label{tab9}
\small
\setlength{\tabcolsep}{3pt}
\begin{tabular}{|l|c|c|c|c|c|}
\hline
\textbf{Model} & \textbf{MAE} & \textbf{MSE} & {\boldmath $R^{2}$} & {\boldmath $T_0$} & {\boldmath $R_0$} \\
\hline
Decision Tree          &  \textbf{19.92} &  \textbf{19{,}907} & \textbf{0.563} & 90      &   1 \\
\hline
Random Forest          &  24.05 &  43{,}295 & 0.049 & 1{,}150    &   0.08 \\
\hline
SVR\tnote{a}           &  45.47 &  30{,}769 & 0.324 & 579{,}800  &  0.0 \\
\hline
Linear Regression      &  64.84 &  44{,}656 & 0.019 & 130     &  0.72 \\
\hline
% Gaussian Process     &  75.15 &  57909 & $-0.18$ &  -      &   - \\
% \hline
PLS\tnote{b}  &  84.90 &  30{,}510 & 0.33 & 130     &   0.693 \\
\hline
SGD\tnote{c}  & \multicolumn{3}{c|}{diverged\tnote{d}} & \textbf{90} & \textbf{1.06} \\
\hline
\end{tabular}
\begin{tablenotes}[flushleft]\footnotesize
\item $T_0$: Average inference time per certificate (nanoseconds) on the 100{,}000 certificate test set.
\item $R_0$: Relative speed compared to Decision Tree (higher is better).
\item[a] SVR: Support Vector Regression.
\item[b] PLS: Partial Least Squares
\item[c] SGD: Stochastic Gradient Descent Regressor.
\item[d] SGD failed to converge: MAE $4.1\times10^{26}$, MSE $1.7\times10^{53}$, $R^2$ $-2.7\times10^{48}$.
\end{tablenotes}
\end{threeparttable}
\end{table}

These models helped to validate our basic approach and benchmark performance expectations, but did not generalize well to the full 201{,}976 certificate test set due to overfitting to the small 20K training set.

% \begin{table}[h]
% \centering
% \begin{threeparttable}
% \caption{Baseline 20K Model Results on Full Test Set}
% \label{tab16}
% \begin{tabularx}{\linewidth}{|l|c|c|c|c|c|}
% \hline
% \textbf{Model} & \textbf{MAE} & \textbf{MSE} & {\boldmath $R^{2}$} & {\boldmath $T_0$} & {\boldmath $R_0$} \\
% \hline
% SVR                      &  433.789 &  204006.9 & -0.013 & 12.337 &   1 \\
% \hline
% Decision Tree                    &  436.39 &  392673.8 & -0.95 &  0.017 &   730.581 \\
% \hline
% Random Forest             &  440.558 &  421622.25 & -1.093 &  0.171 &  72.34 \\
% \hline
% Linear Regression               &  475.39 &  439881.375 & -1.184 &   0.016 &  764.062 \\
% \hline
% Partial Least Squares               &  484.582 &  422403.9 & -1.097 &  0.025 &   491.691 \\
% \hline
% SGD Regressor                 &  4.2E26 &  inf & -inf &  0.015 &   849.848 \\
% \hline
% \end{tabular}
% \begin{tablenotes}[flushleft]\footnotesize
% \item $T_0$: Inference time on the 100,000 certificates in seconds.
% \item $R_0$: Relative inference time to the best model (Decision Tree).
% \item $^*$ SGD Regressor: Stochastic Gradient Descent Regressor. 
% \end{tablenotes}
% \end{threeparttable}
% \end{table}

\subsection{Procedure}
After establishing baseline performance, we followed a systematic procedure to identify and optimize high-performing models:
\begin{enumerate}
  \item \textbf{Model selection:} We trained 25 models on the full training set of 807{,}903 certificates, including ensemble methods 
  (Random Forest, Extra Trees) \cite{pedregosa2011scikit}, gradient boosting frameworks (XGBoost \cite{chen2016xgboost}, LightGBM \cite{ke2017lightgbm}, CatBoost \cite{prokhorenkova2018catboost}), 
  and neural networks \cite{paszke2019pytorch}.
  
  \item \textbf{Candidate filtering:} We selected the top 11 models with MAE $< 11.4$ and $R^2 > 0.89$ for further optimization.
  
  \item \textbf{Feature engineering:} We applied additional preprocessing techniques and engineered new features to improve 
  prediction performance.
  
  \item \textbf{Random seed optimization:} We trained the top six models 50 times each with different random seeds (0-49) 
  to quantify performance variability due to stochastic initialization and identify optimal seeds for these models.
  
  \item \textbf{Hyperparameter tuning:} We performed Bayesian optimization over hyperparameter spaces for each of the top models 
  using their optimal random seeds to identify best configurations.
  
  \item \textbf{Results and analysis:} We presented the tuned model results on the full test and post-deployment sets. Then, we ran an analysis on model performance, residual distributions, tradeoffs between test MAE and speed, severity-tier classification accuracy, and ranking quality (NDCG@$k$).

  \item \textbf{Feature importance analysis:} We analyzed which certificate attributes contributed most to predictions in
  our best-performing models.
\end{enumerate}

\subsubsection{Model Selection}

We trained 25 diverse regression models to identify high-performing architectures before investing in optimization. 
Since training on our dataset required minimal computational resources for most models, we empirically evaluated this 
broad set rather than pre-selecting based on assumptions. We used default hyperparameters for each model to establish 
baseline performance.
% and removed models that did not scale to the full dataset size (e.g., Gaussian Process Regressor, 
% which has $O(n^2)$ space complexity).

We evaluated models from multiple families:

\begin{itemize}
    \item \textbf{Tree-based models} \cite{pedregosa2011scikit}: Decision Tree, Random Forest, Extra Trees, Gradient Boosting, 
    Hist Gradient Boosting, AdaBoost
    \item \textbf{Gradient boosting frameworks}: LightGBM \cite{ke2017lightgbm}, CatBoost \cite{prokhorenkova2018catboost}, XGBoost \cite{chen2016xgboost}
    \item \textbf{Linear models} \cite{pedregosa2011scikit}: Ridge, RidgeCV, Lasso, LassoLars, Elastic Net, ElasticNetCV, Bayesian Ridge, 
    ARD, Huber, Tweedie, Passive Aggressive
    \item \textbf{Neural networks}: MLP \cite{pedregosa2011scikit}, PyTorch Neural Networks \cite{paszke2019pytorch}
    \item \textbf{Other}: KNeighbors \cite{pedregosa2011scikit}, SVR \cite{pedregosa2011scikit}
\end{itemize}

After initial evaluation, we created ensemble models by combining the top three performers by MSE (Extra Trees, Random Forest, 
and CatBoost):

\begin{itemize}
    \item Voting Regressor: averages predictions from Extra Trees, Random Forest, and CatBoost
    \item Stacking Regressor: uses a meta-learner (RidgeCV) to combine Extra Trees, Random Forest, and CatBoost predictions
\end{itemize}

Table~\ref{tab5} presents model performance on the test set, including MAE, MSE, $R^2$ score, average inference time per 
certificate ($T_0$, in nanoseconds), and relative speed compared to Extra Trees ($R_0$). To measure production-relevant 
inference times, we converted each model to ONNX format and used ONNXRuntime for prediction. We report the average 
inference time across 10 runs on the full test set of 201{,}976 certificates, with $R_0$ calculated as the Extra Trees 
inference time (2114 nanoseconds) divided by each model's inference time.

Extra Trees achieved the best overall performance with MAE of 6.593, MSE of 2221.819, and $R^2$ of 0.929. However, 
Decision Tree offered the best performance-speed tradeoff: 32.9$\times$ faster than Extra Trees while achieving competitive 
performance (MAE of 6.72, MSE of 2415.246, $R^2$ of 0.923).

\begin{table}[h]
\centering
\begin{threeparttable}
\caption{Stock Model Results}
\label{tab5}
\begin{tabular}{|@{\hspace{2pt}}l@{\hspace{2pt}}|c|c|c|c|c|}
\hline
\textbf{Model} & \textbf{MAE} & \textbf{MSE} & {\boldmath $R^{2}$} & {\boldmath $T_0$} & {\boldmath $R_0$} \\
\hline
Extra Trees                      &  \textbf{6.593} &  2221.81 & 0.929 &  2114   &  1.000 \\
\hline
Random Forest                    &  6.694 &  2235.67 & 0.929 &  1728   &  1.223 \\
\hline
Decision Tree                    &  6.720 &  2415.25 & 0.923 &  \textbf{64.36}  &  \textbf{32.846} \\
\hline
Stacking Regressor               &  6.972 &  \textbf{2203.97} & \textbf{0.930} &  5798   &  0.365 \\
\hline
Voting Regressor                 &  7.361 &  2236.60 & 0.929 &  5996   &  0.353 \\
\hline
CatBoost                         &  7.470 &  2271.89 & 0.928 &  237.7  &  8.894 \\
\hline
XGBoost                          & 7.666 &   2390.48 & 0.924 &  696.5  &  3.035 \\
\hline
LightGBM                        &  8.267 &  2426.38 & 0.923 &  623.8  &  3.389 \\
\hline
Hist Gradient                    &  8.548 &  2488.49 & 0.921 &  623.8  &  3.389 \\
\hline
MLP                              &  9.214 &  2658.88 & 0.915 &  356.5  &  5.930 \\
\hline
Gradient Boosting                & 11.380 &  3343.62 & 0.894 &  371.3  &  5.694 \\
\hline
ARD                              & 20.918 &  6822.29 & 0.783 &  69.32  &  30.496 \\
\hline
RidgeCV                          & 22.454 &  6795.04 & 0.784 &  84.17  &  25.116 \\
\hline
Bayesian Ridge                    & 22.457 &  6795.07 & 0.784 &  69.32  &  30.496 \\
\hline
Linear Regression                & 22.457 &  6795.20 & 0.784 &  79.22  &  26.685 \\
\hline
Ridge                            & 22.457 &  6793.10 & 0.784 &  84.17  &  25.116 \\
\hline
Lasso                            & 24.684 &  8897.72 & 0.717 &  74.27  &  28.464 \\
\hline
LassoLars                        & 24.684 &  8892.72 & 0.717 &  133.7  &  15.812 \\
\hline
Huber                            & 29.279 & 16525.6 & 0.475 &  84.17  &  25.116 \\
\hline
AdaBoost                         & 31.235 &  9002.41 & 0.714 &  346.6  &  6.099 \\
\hline
Elastic Net                       & 36.214 & 14615.4 & 0.535 &  \textbf{64.36}  &  \textbf{32.846} \\
\hline
Tweedie                          & 37.036 & 15103.2 & 0.520 &  193.1  &  10.948 \\
\hline
Elastic NetCV                     & 39.813 & 15785.1 & 0.498 &  69.32  &  30.496 \\
\hline
Passive Aggressive               & 57.379 & 16690.1 & 0.469 &  89.12  &  23.721 \\
\hline
SVR                              & 433.79 & 204006 & -0.01 &  61540  &  0.034 \\
\hline
\end{tabular}
\begin{tablenotes}[flushleft]\footnotesize
\item $T_0$: Average inference time per certificate (nanoseconds) on the test set of 201{,}976 certificates.
\item $R_0$: Relative speed compared to Extra Trees (higher is better).
\end{tablenotes}
\end{threeparttable}
\end{table}

\subsubsection{Feature Engineering}\label{sec2}
After identifying the top 11 models from our initial survey, we systematically engineered new features to improve prediction performance. 
As shown in Table~\ref{tab5}, these models achieved MAE $< 11.4$, MSE $< 3344$, and $R^2 > 0.89$:
\begin{itemize}
    \item \textbf{Extra Trees} \cite{pedregosa2011scikit}
    \item \textbf{Random Forest} \cite{pedregosa2011scikit}
    \item \textbf{Decision Tree} \cite{pedregosa2011scikit}
    \item \textbf{Stacking Regressor} (Extra Trees, Random Forest, and CatBoost) \cite{pedregosa2011scikit}
    \item \textbf{Voting Regressor} (Extra Trees, Random Forest, and CatBoost) \cite{pedregosa2011scikit}
    \item \textbf{CatBoost} \cite{prokhorenkova2018catboost}
    \item \textbf{XGBoost} \cite{chen2016xgboost}
    \item \textbf{LightGBM} \cite{ke2017lightgbm}
    \item \textbf{Hist Gradient Boosting} \cite{pedregosa2011scikit}
    \item \textbf{MLP} \cite{pedregosa2011scikit}
    \item \textbf{Gradient Boosting} \cite{pedregosa2011scikit}
\end{itemize}

\begin{figure}[h]
  \centering
  \includegraphics[width=0.8\linewidth]{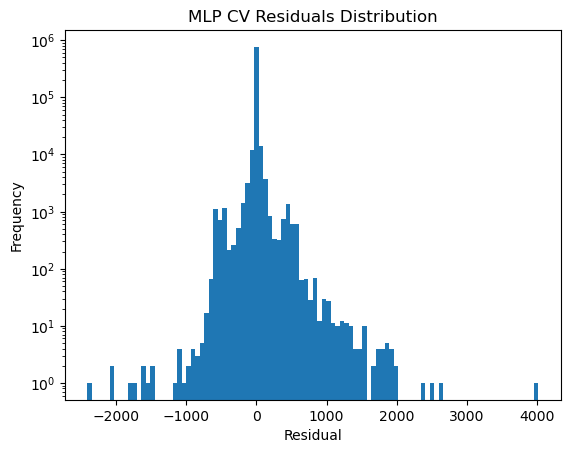}
  \caption{MLP cross-validation residuals $r = y_t - \hat{y}$ across all five folds show systematic errors at $\pm$500, indicating a missing feature for negative serial numbers.}
  \label{fig6}
\end{figure}

\textbf{Residual Analysis and Initial Feature Discovery.}
We began by analyzing cross-validation residuals from the baseline models trained on the original 69 features. 
As shown in Fig.~\ref{fig6}, the residual distribution for MLP revealed distinct peaks at approximately $\pm$500 risk score units. 
This pattern suggested a systematic prediction error tied to a specific defect with a 1000-point penalty in our risk scoring system. 
Upon investigation, we identified that certificates with negative serial numbers accounted for most of these residuals.
A negative serial number constitutes an egregious violation of RFC 5280 \cite{cooper2008rfc5280}
and was scored at 1000 points due to the strong indication that the CA is not built on standard, vetted enterprise software. Because affected issuers produced negative serial numbers in approximately half of their certificates, this defect appeared stochastically and could not be predicted from other certificate attributes.
Adding a binary ``has negative serial number'' indicator dramatically improved performance across 
all models, as evidenced by the transition from $F_0$ to $F_1$ in Table~\ref{tab17}.

\textbf{Systematic Feature Engineering.}
Beyond the negative serial number indicator, we engineered features across four categories:

\textit{Temporal features:} We converted validity period from months to days for finer granularity, computed as 
(``notAfter'' - ``notBefore'') in days rather than months.

\textit{Wildcard indicators:} We created three boolean features to capture wildcard usage patterns in Subject Alternative Names: 
``has wildcard'' (any wildcard present), ``has one wildcard'' (exactly one), and ``has several wildcards'' (multiple wildcards).

\textit{Extended Key Usage (EKU) indicators:} We added five binary features for common EKU object identifiers:
\begin{itemize}
    \item Server Authentication: 1.3.6.1.5.5.7.3.1
    \item Client Authentication: 1.3.6.1.5.5.7.3.2
    \item Code Signing: 1.3.6.1.5.5.7.3.3
    \item Secure Email: 1.3.6.1.5.5.7.3.4
    \item IP Security End System: 1.3.6.1.5.5.7.3.5
\end{itemize}

\textit{Issuer geography and SAN details ($F_4$ only):} We added issuer country indicator variables for the 14 most common
countries among our certificate inventory (US, GB, BE, AT, CA, FR, UK, NL, NA (Namibia), AU, NO, CN, JP, IN) plus indicators for empty and other countries. 
We also decomposed the total SAN count into type-specific counts: DNS names, IP addresses, email addresses, URIs, 
other names, directory names, EDI party names, registered IDs, UPNs, and DS replication identifiers. Finally, we 
added serial number length and signing algorithm as additional structural features.

\textbf{Feature Set Evaluation.}
Table~\ref{tab17} presents five-fold cross-validation (CV) MAE results for five feature configurations:

\begin{itemize}
    \item $F_0$: Baseline 69 features
    \item $F_1$: $F_0$ + negative serial number indicator + validity period in days (replacing months)
    \item $F_2$: $F_1$ + wildcard indicators
    \item $F_3$: $F_1$ + EKU indicators
    \item $F_4$: $F_3$ + wildcard indicators + issuer country indicators + SAN type counts + serial number length + signing algorithm
\end{itemize}

The negative serial number feature in $F_1$ produced dramatic improvements across all models, with MAE reductions 
ranging from 58\% (Extra Trees: 6.66 $\rightarrow$ 2.77) to 67\% (Gradient Boosting: 11.38 $\rightarrow$ 3.78). 
Wildcard features ($F_2$) provided marginal gains, with the largest impact on MLP (5.28 $\rightarrow$ 5.09). 
EKU indicators ($F_3$) improved performance for all models. The comprehensive feature 
set $F_4$ achieved the best performance for 10 of 11 models, with only MLP performing slightly better on $F_3$. 
Table~\ref{tab17} highlights the lowest CV MAE score for each model in green.

\begin{table}[h]
\centering
\begin{threeparttable}
\caption{Feature Engineering CV MAE Results}
\label{tab17}
\begin{tabular}{|l|c|c|c|c|c|}
\hline
\textbf{Model} & {\boldmath $F_0$} & {\boldmath $F_1$} & {\boldmath $F_2$} & {\boldmath $F_3$} & {\boldmath $F_4$} \\
\hline
Extra Trees                  & 6.66 & 2.77 & 2.66 & 2.55 & \cellcolor{medgreen}2.31 \\
\hline
Random Forest                & 6.76 & 2.88 & 2.76 & 2.65 & \cellcolor{medgreen}2.41 \\
\hline
Decision Tree                & 6.73 & 2.85 & 2.74 & 2.63 & \cellcolor{medgreen}2.40 \\
\hline
Stacking Regressor           & 6.93 & 2.97 & 2.90 & 2.83 & \cellcolor{medgreen}2.58 \\
\hline
Voting Regressor             & 6.93 & 3.03 & 2.92 & 2.80 & \cellcolor{medgreen}2.57 \\
\hline
CatBoost                     & 7.50 & 3.58 & 3.46 & 3.33 & \cellcolor{medgreen}3.15 \\
\hline
LightGBM                     & 7.63 & 4.36 & 4.29 & 4.15 & \cellcolor{medgreen}4.00 \\
\hline
Hist Gradient Boosting      & 8.22 & 4.65 & 4.54 & 4.43 & \cellcolor{medgreen}4.18 \\
\hline
MLP                          & 8.44 & 5.28 & 5.09 & \cellcolor{medgreen}4.65 & 4.68 \\
\hline
XGBoost                      & 9.32 & 3.78 & 3.55 & 3.40 & \cellcolor{medgreen}3.40 \\
\hline
Gradient Boosting            & 11.38 & 3.78 & 3.56 & 3.40 & \cellcolor{medgreen}3.28 \\
\hline
\end{tabular}
\begin{tablenotes}[flushleft]\footnotesize
\item $F_0$: Original 69 features.
\item $F_1$: $F_0$ + negative serial number indicator + validity period in days (replacing months).
\item $F_2$: $F_1$ + wildcard indicators (has wildcard, has one wildcard, has several wildcards).
\item $F_3$: $F_1$ + EKU indicators (5 common usage types).
\item $F_4$: $F_3$ + wildcard indicators + issuer country indicators (16 categories) + SAN type counts (10 types) + serial length + signing algorithm.
\end{tablenotes}
\end{threeparttable}
\end{table}

\subsubsection{Random Seed Optimization}

After feature engineering, we investigated whether model performance could be further improved through random seed optimization.
Many tree-based and gradient boosting models incorporate stochastic elements during training (e.g., bootstrap sampling in Random
Forest, random feature selection, random row sampling). The random seed controls these stochastic processes, and different seeds
can lead to different model initializations and training paths, potentially affecting final performance.

We evaluated the top six models that could be trained in a reasonable amount of time based on their CV MAE performance from feature engineering: Decision Tree, Random Forest,
CatBoost, LightGBM, Hist Gradient Boosting, and Gradient Boosting. For each model, we trained 50
iterations with different random seeds (0-49), holding all other hyperparameters at their default values. We used each model's
optimal feature set from Table~\ref{tab17} and recorded the five-fold cross-validation MAE for each seed. This quantified the
variability due to random initialization and surfaced seeds that consistently performed better. For Gradient Boosting, we observed
negligible seed sensitivity and include its statistics for completeness.

\begin{figure}[h]
\centering
\begin{tikzpicture}
\begin{axis}[
width=0.95\linewidth,
height=8cm,
ylabel={5-Fold CV MAE},
xlabel={Model},
boxplot/draw direction=y,
xtick={1,2,3,4,5,6},
xticklabels={Decision Tree, Random Forest, CatBoost, LightGBM, Hist GB, Gradient Boosting},
x tick label style={rotate=45, anchor=east, font=\small},
ymin=2.35,
ymax=8.1,
ytick distance=0.5,
minor y tick num=3,
grid=both,
grid style={dashed, gray!30},
minor grid style={dotted, gray!20},
boxplot={
box extend=0.5,
},
ylabel style={font=\large},
xlabel style={font=\large},
tick label style={font=\normalsize}
]

% Decision Tree (position 1)
\addplot+[
  boxplot prepared={
    lower whisker=2.380000,
    lower quartile=2.398632,
    median=2.402379,
    upper quartile=2.406991,
    upper whisker=2.416000,
    average=2.402277
  },
  fill=blue!30,
  draw=blue!70!black,
  line width=1pt
] coordinates {};

% Random Forest (position 2) -- updated tight range 2.409--2.410
\addplot+[
  boxplot prepared={
    lower whisker=2.409000,
    lower quartile=2.409500,
    median=2.410000,
    upper quartile=2.410000,
    upper whisker=2.410000,
    average=2.409500
  },
  fill=green!30,
  draw=green!70!black,
  line width=1pt
] coordinates {};

% CatBoost (position 3)
\addplot+[
  boxplot prepared={
    lower whisker=3.125000,
    lower quartile=3.140436,
    median=3.148363,
    upper quartile=3.153941,
    upper whisker=3.168000,
    average=3.147093
  },
  fill=orange!30,
  draw=orange!70!black,
  line width=1pt
] coordinates {};

% LightGBM (position 4)
\addplot+[
  boxplot prepared={
    lower whisker=3.977000,
    lower quartile=3.991951,
    median=4.001878,
    upper quartile=4.007439,
    upper whisker=4.039000,
    average=4.000483
  },
  fill=red!30,
  draw=red!70!black,
  line width=1pt
] coordinates {};

% Hist Gradient Boosting (position 5)
\addplot+[
  boxplot prepared={
    lower whisker=4.163000,
    lower quartile=4.189162,
    median=4.204148,
    upper quartile=4.216832,
    upper whisker=4.244000,
    average=4.204520
  },
  fill=purple!30,
  draw=purple!70!black,
  line width=1pt
] coordinates {};

% Gradient Boosting (position 6) -- near-zero variance around ~7.934
\addplot+[
  boxplot prepared={
    lower whisker=7.933000,
    lower quartile=7.933500,
    median=7.933900,
    upper quartile=7.934000,
    upper whisker=7.934000,
    average=7.933900
  },
  fill=gray!30,
  draw=gray!70!black,
  line width=1pt
] coordinates {};

% Annotate each box with mean +/- std
\node[above, font=\scriptsize, align=center] at (axis cs:1,2.44)
  {$\bar{x}$=2.402\\$\sigma$=0.008};
\node[above, font=\scriptsize, align=center] at (axis cs:2,2.44)
  {$\bar{x}$=2.410\\$\sigma$\,$<$\,0.001};
\node[above, font=\scriptsize, align=center] at (axis cs:3,3.20)
  {$\bar{x}$=3.147\\$\sigma$=0.011};
\node[above, font=\scriptsize, align=center] at (axis cs:4,4.07)
  {$\bar{x}$=4.000\\$\sigma$=0.012};
\node[above, font=\scriptsize, align=center] at (axis cs:5,4.28)
  {$\bar{x}$=4.205\\$\sigma$=0.020};
\node[above, font=\scriptsize, align=center] at (axis cs:6,7.96)
  {$\bar{x}$=7.934\\$\sigma$\,$<$\,0.001};

\end{axis}

\end{tikzpicture}
\caption{Distribution of 5-fold CV MAE across random seeds for six models. Boxes show the interquartile range (Q1--Q3),
the horizontal line is the median, whiskers span min to max, and annotations show the mean ($\bar{x}$) and standard deviation ($\sigma$) for each model. All MAE values are from 5-fold cross-validation.}
\label{fig:seed_opt}
\label{tab:seed_stats}
\end{figure}
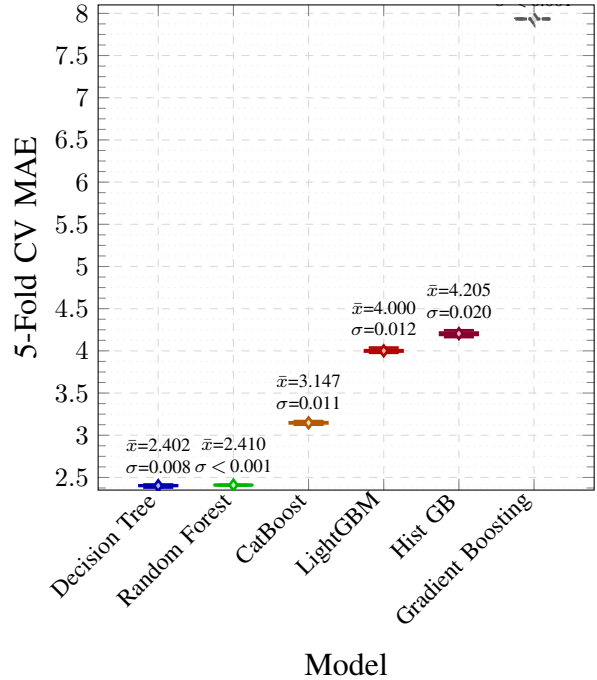

Fig.~\ref{fig:seed_opt} presents the distribution and summary statistics of CV MAE scores across random seeds for the six models.

The results reveal varying degrees of seed sensitivity. Random Forest shows an extremely tight range (2.409-2.410 MAE), which 
reflects strong stability from ensemble averaging. Decision Tree varies slightly more (2.380-2.416 MAE), which is expected
without ensemble effects. Among the gradient boosting models, CatBoost and LightGBM exhibit modest variance, while Hist Gradient
Boosting is the most sensitive (4.163-4.244 MAE). Gradient Boosting shows near-zero variance (7.933-7.934 MAE) and is included
to document its insensitivity to seed choice.

For each model, we identified the top three seeds with the lowest CV MAE and carried those forward as hyperparameters into the
Bayesian optimization phase:

\begin{itemize}
\item \textbf{Decision Tree}: seeds 12, 35, 7 %(MAE range: 2.380-2.416)
\item \textbf{Random Forest}: seeds 48, 46, 45 %(MAE range: 2.409-2.410)
\item \textbf{CatBoost}: seeds 32, 39, 19 %(MAE range: 3.125-3.168)
\item \textbf{LightGBM}: seeds 12, 40, 44 %(MAE range: 3.977-4.039)
\item \textbf{Hist Gradient Boosting}: seeds 41, 39, 48 %(MAE range: 4.163-4.244)
\item \textbf{Gradient Boosting}: seeds 1, 49, 23 %(MAE range: 7.933-7.934)
\end{itemize}

For the remaining models (Extra Trees, XGBoost, MLP, and the ensemble models), we used default random seed values during
hyperparameter tuning, and added three arbitrary seeds (1, 2, and 3) to introduce light stochasticity and verify stability at minimal cost. While seed optimization offered modest gains (typically 0.5-2\% MAE reduction), it is a low-effort
lever with no additional feature engineering. We recommend evaluating 3--5 seeds for seed-sensitive models (e.g., Hist Gradient Boosting) and defaulting to stable ensembles (e.g., Random Forest).

\subsubsection{Hyperparameter Tuning}

After identifying the best feature set and optimal random seeds for each model, we performed hyperparameter tuning on the nine non-ensemble models. 
We focused on the non-ensemble models first, because we wanted to find the best hyperparameters for each base estimator before tuning 
the voting and stacking ensembles. For each model, we selected the most important four to six hyperparameters to reduce the 
search space and find optimal configurations more efficiently (all unset parameters were set to their default values).

We utilized Bayesian optimization with five-fold cross-validation \cite{akiba2019optuna} to perform the hyperparameter search. For the six 
models that underwent random seed optimization (Decision Tree, Random Forest, CatBoost, LightGBM, Hist Gradient Boosting, and Gradient Boosting), 
we included the top three performing seeds and the seed 0 as hyperparameters in the search space. This allowed the optimization process to 
explore configurations with the most promising random initializations alongside other hyperparameters.

As indicated in Table~\ref{tab19}, hyperparameter tuning produced substantial improvements for most models. LightGBM showed the most dramatic improvement, 
with CV MAE decreasing from 3.977 to 2.42--a 39.15\% reduction. Gradient Boosting achieved the second-largest improvement, with MAE decreasing 
from 3.28 to 2.01--a 38.72\% reduction. This improvement was primarily driven by increasing n\_estimators from the default value of 100 to 800, 
allowing the model to fit the training data much more effectively. Hist Gradient Boosting also achieved significant gains, with an MAE reduction 
of 32.50\%, while XGBoost improved by 26.18\%. CatBoost showed a substantial 23.84\% reduction in MAE, from 3.125 to 2.38. MLP showed notable improvement 
as well, decreasing from 4.68 to 3.62 MAE--a 22.65\% reduction achieved through optimization of the network architecture and learning parameters. 
These gradient boosting models and neural networks benefited from both architectural complexity and optimized learning rates.

Tree-based ensemble models showed more modest improvements. Extra Trees maintained its performance at 2.31 MAE, suggesting its default parameters 
were already well-suited to this problem. Random Forest showed minimal improvement from 2.409 to 2.38 MAE (1.20\% reduction). Decision Tree 
maintained its performance at 2.38 MAE, as the default configuration--which allows trees to grow until all leaves are pure or contain a single 
sample--was already optimal for this dataset, and the hyperparameter search confirmed these default settings.
The optimal hyperparameters identified were:

\begin{itemize}
    \item \textbf{Extra Trees} \cite{pedregosa2011scikit}: n\_estimators: 700, max\_features: 0.9, min\_samples\_leaf: 1, random\_state: 0
    \item \textbf{Random Forest} \cite{pedregosa2011scikit}: n\_estimators: 361, max\_features: 0.6, min\_samples\_leaf: 1, random\_state: 24
    \item \textbf{Decision Tree} \cite{pedregosa2011scikit}: min\_samples\_leaf: 1, ccp\_alpha: 0.0, criterion: squared\_error, random\_state: 12
    \item \textbf{CatBoost} \cite{prokhorenkova2018catboost}: learning\_rate: 0.167, n\_estimators: 3800, l2\_leaf\_reg: 1.121, depth: 10, random\_state: 39
    \item \textbf{LightGBM} \cite{ke2017lightgbm}: learning\_rate: 0.181, n\_estimators: 2000, min\_child\_samples: 10, subsample: 0.890, colsample\_bytree: 0.772, num\_leaves: 127, random\_state: 12
    \item \textbf{Hist Gradient Boosting} \cite{pedregosa2011scikit}: learning\_rate: 0.154, max\_iter: 500, min\_samples\_leaf: 6, max\_bins: 255, max\_leaf\_nodes: 255, l2\_regularization: 0.0024, random\_state: 41
    \item \textbf{MLP} \cite{pedregosa2011scikit}: hidden\_layer\_sizes: (512,), max\_iter: 700, solver: lbfgs, learning\_rate: constant, learning\_rate\_init: 0.171, random\_state: 1
    \item \textbf{Gradient Boosting} \cite{pedregosa2011scikit}: learning\_rate: 0.139, n\_estimators: 800, min\_samples\_leaf: 3, max\_depth: 12, subsample: 0.834, max\_features: 0.7, alpha: 0.852, random\_state: 49
    \item \textbf{XGBoost} \cite{chen2016xgboost}: learning\_rate: 0.3, min\_child\_weight: 1.0, subsample: 1.0, colsample\_bytree: 1.0, n\_estimators: 1800, random\_state: 0
\end{itemize}

\begin{table}[h]
\centering
\begin{threeparttable}
\caption{Hyperparameter CV Results}
\label{tab19}
\begin{tabular}{|l|c|c|c|}
\hline
\textbf{Model} & \textbf{MAE}{\boldmath$_{0}$} & \textbf{MAE}{\boldmath$_1$} & \textbf{Impr. (\%)} \\
\hline
Extra Trees                     & 2.31 &  2.31 & 0.00 \\
\hline
Random Forest                  & 2.409 &  2.38 & 1.20 \\
\hline
Decision Tree                  & 2.380 &  2.38 & 0.00 \\
\hline
CatBoost               & 3.125 &  2.38 & 23.84 \\
\hline
XGBoost             & 3.40 &  2.51 & 26.18 \\
\hline
LightGBM             & 3.977 &  2.42 & 39.15 \\
\hline
Hist Gradient Boosting            & 4.163 &  2.81 & 32.50 \\
\hline
MLP            & 4.68 &  3.62 & 22.65 \\
\hline
Gradient Boosting         & 3.28 &  2.01 & 38.72 \\
\hline
\end{tabular}
\begin{tablenotes}[flushleft]\footnotesize
\item MAE$_0$: Best CV MAE after feature engineering and random seed optimization (minimum from Fig.~\ref{fig:seed_opt} for optimized models, Table~\ref{tab17} for others).
\item MAE$_1$: Best CV MAE after hyperparameter optimization with Bayesian search.
\item Impr. (\%): Percentage improvement calculated as $\frac{\text{MAE}_0 - \text{MAE}_1}{\text{MAE}_0} \times 100$.
\end{tablenotes}
\end{threeparttable}
\end{table}

After discovering these optimal hyperparameters for the base models, we tuned the voting and stacking ensemble regressors. 
The ensemble models use Extra Trees, Random Forest, and CatBoost as base estimators with their newly optimized hyperparameters. 
We found the following optimal ensemble configurations:

\begin{itemize}
    \item \textbf{Stacking Regressor} \cite{pedregosa2011scikit}: ridge\_alpha:~0.220
    \item \textbf{Voting Regressor} \cite{pedregosa2011scikit}: weights: [0.295, 0.384, 0.321]
\end{itemize}

Ensemble-specific hyperparameter tuning yielded modest but meaningful performance gains, with the Voting Regressor achieving a 9.72\% 
improvement in MAE compared to 8.53\% for the Stacking Regressor, demonstrating that optimizing ensemble-level parameters 
beyond base estimator tuning provides additional predictive value.

\begin{table}[h]
\centering
\begin{threeparttable}
\caption{Hyperparameter Ensemble CV Results}
\label{tab20}
\begin{tabular}{|l|c|c|c|}
\hline
\textbf{Model} & \textbf{MAE}{\boldmath$_{0}$} & \textbf{MAE}{\boldmath$_1$} & \textbf{Impr. (\%)} \\
\hline
Stacking Regressor              & 2.58 & 2.36 & 8.53 \\
\hline
Voting Regressor              & 2.57 & 2.32 & 9.72 \\
\hline
\end{tabular}
\begin{tablenotes}[flushleft]\footnotesize
\item MAE$_0$: CV MAE with tuned base estimators but default ensemble parameters (from Table~\ref{tab17}).
\item MAE$_1$: CV MAE after ensemble-specific hyperparameter tuning.
\item Impr. (\%): Percentage improvement calculated as $\frac{\text{MAE}_0 - \text{MAE}_1}{\text{MAE}_0} \times 100$.
\end{tablenotes}
\end{threeparttable}
\end{table}

\subsection{Hardware and Runtime}
We trained and tested our models on the hardware described in Table~\ref{tab6}. 

\begin{table}[h]
  \caption{Hardware Specs}
  \centering
  \label{tab6}
  \begin{tabular}{|l|r|}
    \hline
    \textbf{Computer}               & \textbf{Dell XPS 9520} \\
    \hline
    CPU & Intel Core i7-12700H 2.3 GHz \\
    \hline
    GPU & NVIDIA RTX 3050 Mobile \\
    VRAM & 4 GB \\
    \hline
    RAM & 32 GB \\
    \hline
    SSD & 477 GB \\
    \hline
    OS & Windows 11 Enterprise \\
    \hline
  \end{tabular}
\end{table}

\subsection{Results}

After systematic optimization through feature engineering, random seed selection, and hyperparameter tuning, we retrained all 11 models on the full 807{,}903-certificate training set using their optimal configurations and evaluated them on the 201{,}976 certificate test set. Table~\ref{tab21} presents the final performance metrics, revealing substantial improvements over the baseline models in Table~\ref{tab5}.

% \newcolumntype{C}{>{\centering\arraybackslash}X}

% \newcolumntype{C}{>{\centering\arraybackslash}X}

\begin{table*}[t]

\centering

\begin{threeparttable}

\caption{Tuned Model Results on the Test Set and the 2026 Post-Deployment Set}

\label{tab21}

\begin{tabular}{|l|c|c|c|c|c|c|c|c|c|}

\hline

\textbf{Model} & \textbf{MAE}{\boldmath$_T$} & \textbf{MAE}{\boldmath$_P$} & {\boldmath $\Delta_{\mathrm{MAE}}$} & \textbf{MSE}{\boldmath$_T$} & \textbf{MSE}{\boldmath$_P$} & {\boldmath $R^2$}{\boldmath$_T$} & {\boldmath $R^2$}{\boldmath$_P$} & {\boldmath $T_0$} & {\boldmath $R_0$} \\

\hline

Extra Trees & \textbf{2.263} & 6.720 & 2.97 & 219.64 & 536.62 & 0.993 & 0.927 & 30430 & 1.00 \\

\hline

Random Forest & 2.348 & \textbf{4.495} & 1.91 & 216.42 & 432.74 & 0.993 & 0.941 & 13640 & 2.23 \\

\hline

CatBoost & 2.357 & 11.184 & 4.75 & 228.10 & 741.19 & 0.993 & 0.899 & 6765 & 4.50 \\

\hline

Decision Tree & 2.395 & 5.204 & 2.17 & 426.67 & 626.13 & 0.986 & 0.915 & \textbf{270} & \textbf{112.70} \\

\hline

Gradient Boosting & 2.397 & 5.256 & 2.19 & 224.72 & 556.45 & 0.993 & 0.924 & 6820 & 4.46 \\

\hline

LightGBM & 2.415 & 7.369 & 3.05 & 233.42 & 604.15 & 0.993 & 0.918 & 12115 & 2.51 \\

\hline

Voting Regressor & 2.416 & 4.496 & \textbf{1.86} & \textbf{211.70} & \textbf{425.58} & 0.993 & \textbf{0.942} & 28765 & 1.06 \\

\hline

XGBoost & 2.482 & 7.064 & 2.85 & 249.59 & 738.24 & 0.992 & 0.899 & 8655 & 3.52 \\

\hline

Stacking Regressor & 2.497 & 5.143 & 2.06 & 204.86 & 444.09 & 0.993 & 0.939 & 28790 & 1.06 \\

\hline

Hist Gradient Boosting & 2.745 & 13.083 & 4.77 & 259.78 & 1363.24 & 0.992 & 0.814 & 1790 & 17.00 \\

\hline

MLP & 3.476 & 10.497 & 3.02 & 358.19 & 1039.52 & 0.989 & 0.858 & 770 & 39.52 \\

\hline

\end{tabular}

\begin{tablenotes}[flushleft]\footnotesize

\item Rows are ordered by test-set MAE. Subscript $T$ denotes the 201{,}976-certificate held-out test set; subscript $P$ denotes the 571{,}374 certificate post-deployment set collected August 5--9, 2026. Bold marks the best value in each post-deployment column and the fastest inference time.

\item Post-deployment values are for the already-tuned models evaluated as deployed: no model was refit, recalibrated, or re-tuned on the 2026 data.

\item $\Delta_{\mathrm{MAE}}$: ratio MAE$_P$ / MAE$_T$ (lower is better; 1.00 would indicate no degradation).

\item $T_0$: Average inference time per certificate (\textit{nanoseconds}) on the held-out test set. $R_0$: Extra Trees inference time divided by the model's, so higher is faster.

\end{tablenotes}

\end{threeparttable}

\end{table*}

\subsubsection{Performance Analysis}

Our optimization pipeline delivered dramatic improvements across all models. Comparing baseline performance (Table~\ref{tab5}) to optimized results (Table~\ref{tab21}), we observe MAE reductions of 65.7\% (Extra Trees: 6.593 $\rightarrow$ 2.263), 64.9\% (Random Forest: 6.694 $\rightarrow$ 2.348), and 64.4\% (Decision Tree: 6.720 $\rightarrow$ 2.395). The gradient boosting frameworks showed even more pronounced gains: CatBoost improved 68.5\% (7.470 $\rightarrow$ 2.357), LightGBM 70.8\% (8.267 $\rightarrow$ 2.415), and Gradient Boosting 78.9\% (11.380 $\rightarrow$ 2.397). All optimized models now achieve $R^2 \geq 0.986$, with nine models reaching $R^2 \ge 0.992$, indicating near-perfect explanatory power.

The ensemble models justify their architectural complexity. Stacking Regressor achieves the lowest MSE (204.86), effectively combining the strengths of Extra Trees, Random Forest, and CatBoost through its meta-learner (RidgeCV). Voting Regressor follows closely with MSE of 211.70 while maintaining competitive MAE (2.416). Both ensembles demonstrate that sophisticated aggregation strategies can extract additional performance beyond individual base estimators, though at the cost of increased inference latency.

\subsubsection{Residual Distribution Analysis}\label{sec:residuals}
\begin{figure}[h]
  \centering
  \includegraphics[width=0.8\linewidth]{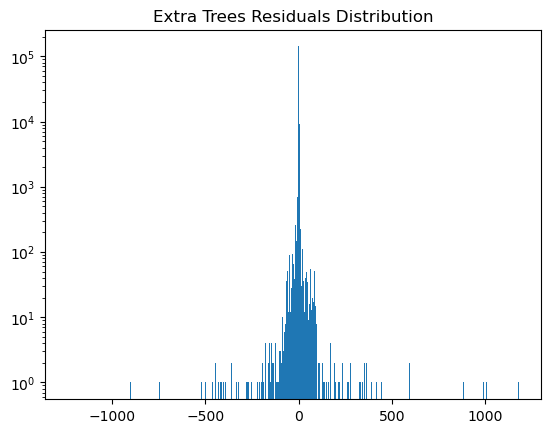}
  \caption{Extra Trees residuals $r = y_t - \hat{y}$}
  \label{fig7}
\end{figure}

\begin{figure}[h]
  \centering
  \includegraphics[width=0.8\linewidth]{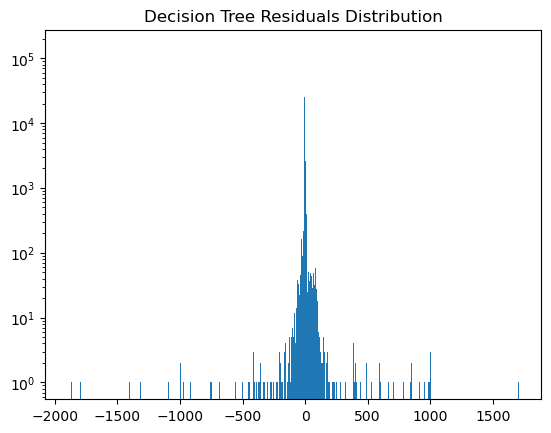}
  \caption{Decision Tree residuals $r = y_t - \hat{y}$}
  \label{fig8}
\end{figure}

Figures~\ref{fig7} and~\ref{fig8} show the residual distributions for the Extra Trees and Decision Tree models, respectively.
Residuals represent the difference between actual and predicted risk scores. A narrow, symmetric distribution centered near zero indicates low bias and good generalization.

The Extra Trees model (Figure~\ref{fig7}) shows a sharply peaked, near-symmetric residual curve, with most errors concentrated around zero. This confirms that the ensemble effectively averages out noise and avoids large over- or under-predictions. The log-scaled y-axis highlights how rarely large deviations occur, suggesting strong error stability across certificate samples.

The Decision Tree model (Figure~\ref{fig8}) exhibits a slightly broader spread compared to the Extra Trees model, which indicates higher variance and occasional overfitting to specific certificate patterns.
However, the Decision Tree model
still generalizes well to the test set.

\subsubsection{Performance-Speed Tradeoffs}\label{sec:tradeoffs}

Decision Tree emerges as the standout model for real-time triage, offering an exceptional balance of accuracy and speed. With an MAE of 2.395 and $R^2$ of 0.986, it sacrifices only 5.9\% MAE performance relative to Extra Trees (2.263) while delivering 113$\times$ faster inference (270ns vs 30{,}430ns per certificate). At this speed, Decision Tree can process approximately 3.7 million certificates per second on a single machine. The question is not whether distributed deterministic analysis can achieve comparable throughput; it can. The question is whether organizations should deploy substantial compute infrastructure to exhaustively analyze millions of compliant certificates when ML triage can identify the subset warranting detailed analysis in seconds on commodity hardware.

For batch triage, where an organization periodically rescores its entire certificate inventory overnight or during a maintenance window, ensemble models offer the highest ranking fidelity. Extra Trees (MAE: 2.263), Random Forest (MAE: 2.348), and CatBoost (MAE: 2.357) form a top tier with nearly indistinguishable performance. CatBoost offers the best compromise within this group, achieving $R^2 = 0.993$ at 4.5$\times$ faster inference than Extra Trees (6{,}765ns vs 30{,}430ns). These models are well-suited for scheduled batch runs where inference latency is less constrained and maximizing ranking quality ensures that downstream deterministic analysis is directed at the certificates most likely to contain critical defects.

Among the gradient boosting frameworks, the performance hierarchy is clear: CatBoost (MAE: 2.357) outperforms LightGBM (2.415), XGBoost (2.482), and Hist Gradient Boosting (2.745), while maintaining competitive inference times. Notably, Gradient Boosting achieves MAE of 2.397--matching Decision Tree's accuracy--but requires 25$\times$ longer inference time (6{,}820ns vs 270ns), demonstrating Decision Tree's efficiency advantage. Hist Gradient Boosting, despite showing 32.5\% improvement during hyperparameter tuning (Table~\ref{tab19}), remains the slowest gradient boosting variant at 2.745 MAE, suggesting diminishing returns from its more complex bin-based splitting strategy.

The MLP neural network (MAE: 3.476, $R^2$: 0.989) lags behind tree-based methods despite architecture and learning parameter optimization. MLP achieves relatively fast inference compared to Extra Trees (770ns, 39.5$\times$ faster than Extra Trees), but it is still 2.85$\times$ slower than Decision Tree (270ns).

\subsubsection{Severity-Tier Classification}\label{sec:tier_eval}

To evaluate whether the regression models support triage, we convert continuous predicted and actual risk scores into severity tiers:
\textit{Low} (score $< 100$), \textit{High Severity} ($100 \leq$ score $< 1000$), and \textit{Critical} (score $\geq 1000$).
While the scoring rubric (Section~3) defines four severity levels, we collapse the O(1) and O(10) tiers into a single Low tier because both receive the same triage action: no escalation to deterministic analysis. This yields three operationally distinct tiers aligned with triage decisions.
Classification accuracy is then computed by comparing the predicted tier to the actual tier for each certificate in the 201{,}976-certificate test set.

\begin{figure}[h]
  \centering
  \includegraphics[width=0.85\linewidth]{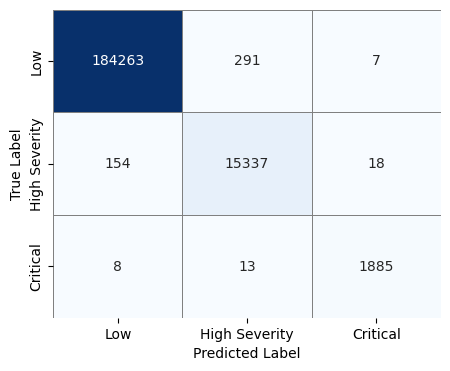}
  \caption{Decision Tree severity-tier confusion matrix on the 201{,}976-certificate test set.}
  \label{fig:cm_dt}
\end{figure}

\begin{figure}[h]
  \centering
  \includegraphics[width=0.85\linewidth]{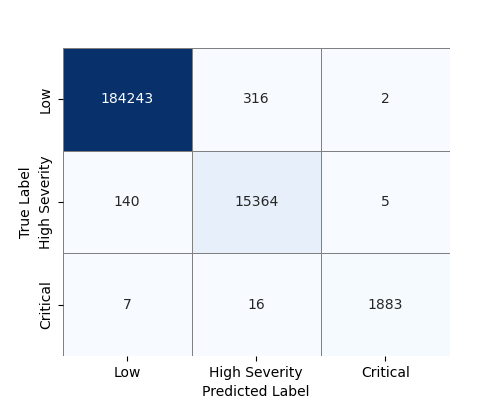}
  \caption{Extra Trees severity-tier confusion matrix on the 201{,}976-certificate test set.}
  \label{fig:cm_et}
\end{figure}

Figures~\ref{fig:cm_dt} and~\ref{fig:cm_et} present the confusion matrices for the Decision Tree and Extra Trees models, respectively.
The Decision Tree correctly classifies 99.76\% of certificates into their true severity tier (201{,}485 of 201{,}976).
Of the 1{,}906 critical certificates, only 21 are misclassified: 8 placed in the Low tier and 13 in high severity.
This yields a critical-tier recall of 98.90\%, meaning virtually all certificates with critical defects (e.g., factorizable RSA keys, missing path length constraints) are surfaced to the top of the triage queue.
The Extra Trees model achieves even higher classification accuracy at 99.76\% (201{,}490 of 201{,}976), with only 23 critical certificates misclassified: 7 placed in Low and 16 in high severity, yielding a critical-tier recall of 98.79\%.

Table~\ref{tab:tier_metrics} reports per-tier precision, recall, and F1 score for the Decision Tree and Extra Trees models.
All three tiers achieve F1 scores above 98\% for both models, confirming that the regression predictions preserve tier boundaries with high fidelity.
The Low tier benefits from its large support (184{,}561 certificates, 91.4\% of the test set), achieving 99.88\% (Decision Tree) and 99.87\% (Extra Trees) F1.
The critical-tier, despite comprising only 0.94\% of the test set (1{,}906 certificates), achieves 98.79\% (Decision Tree) and 99.21\% (Extra Trees) F1, demonstrating that neither model sacrifices minority-class accuracy for majority-class performance.

\begin{table}[h]
\centering
\begin{threeparttable}
\caption{Per-Tier Classification Metrics on the Held-Out Test Set}
\label{tab:tier_metrics}
\footnotesize
\setlength{\tabcolsep}{4pt}
\begin{tabular}{|l|l|c|c|c|}
\hline
\textbf{Model} & \textbf{Tier} & \textbf{Precision} & \textbf{Recall} & \textbf{F1} \\
\hline
Decision Tree & Low       & 99.91\% & 99.84\% & 99.88\% \\
              & High Sev. & 98.06\% & 98.89\% & 98.47\% \\
              & Critical  & 98.69\% & \textbf{98.90\%} & 98.79\% \\
\hline
Extra Trees   & Low       & 99.92\% & 99.83\% & 99.87\% \\
              & High Sev. & 97.88\% & 99.07\% & 98.47\% \\
              & Critical  & \textbf{99.63\%} & 98.79\% & \textbf{99.21\%} \\
\hline
\end{tabular}
\begin{tablenotes}[flushleft]\footnotesize
\item Computed on the 201{,}976-certificate held-out test set using the tier boundaries defined above: Low ($<100$), High Severity ($100$--$999$), Critical ($\geq 1000$).
\item Support is identical across models because tiers derive from ground-truth scores: Low 184{,}561 (91.4\%), High Severity 15{,}509, Critical 1{,}906 (0.94\%).
\item Overall tier accuracy: Decision Tree 99.76\%, Extra Trees 99.76\%.
\end{tablenotes}
\end{threeparttable}
\end{table}

From a practitioner's perspective, both models achieve 99.76\% overall tier accuracy. The Decision Tree's critical recall of 98.90\% means that in a deployment where only the top-ranked certificates receive deterministic analysis, the model would correctly escalate nearly all critical defects. Extra Trees matches this with 98.79\% critical recall while achieving higher critical precision (99.63\% vs.\ 98.69\%), reducing false alarms in the triage queue. The 8 (Decision Tree) and 7 (Extra Trees) critical certificates misclassified as Low represent each model's false-negative floor; these would still be identified in periodic full-inventory scans.

\subsubsection{Ranking Quality Evaluation}\label{sec:ranking}

The triage framing requires that the model not only predict accurate scores, but also produce a correct ranking: certificates with higher true risk must appear earlier in the model's ranked output.

\begin{figure}[h]
  \centering
  \includegraphics[width=0.85\linewidth]{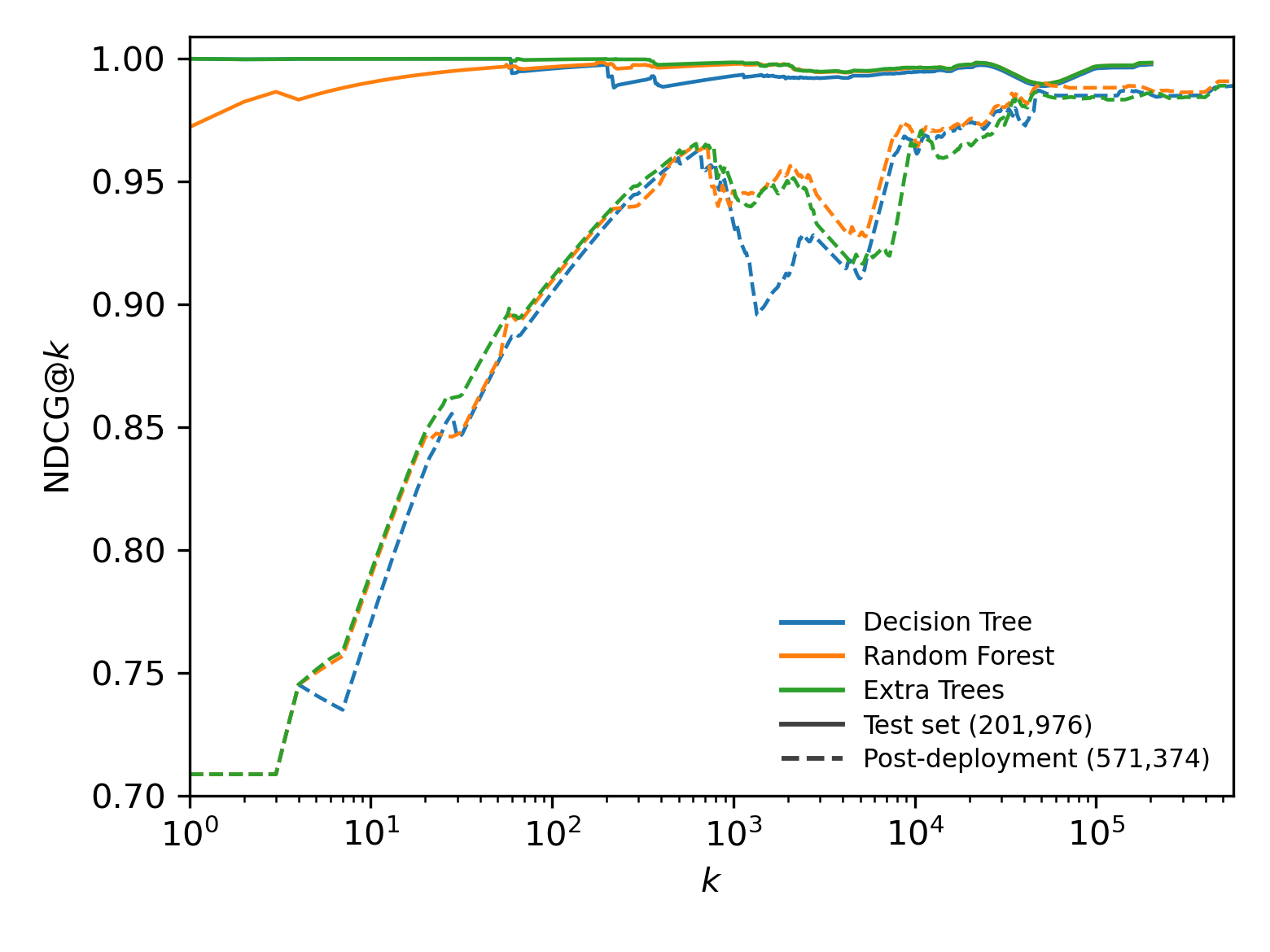}
  \caption{Select models NDCG@$k$ on the 201{,}976-certificate test set and 571{,}374. NDCG@$k \approx 1.0$ for $k \leq 100$ indicates perfect ranking of the highest-risk certificates.}
  \label{fig:ndcg}
\end{figure}

Figure~\ref{fig:ndcg} shows NDCG@$k$ for select models on both evaluation sets. On the 201{,}976 certificate test set, Decision Tree and Extra Trees hold 
NDCG@$k$ above 0.988 and 0.989 respectively at every depth, with aggregate NDCG of 0.9977 and 0.9984, and they confirm strong ranking fidelity across the entire risk spectrum. Random Forest has a slightly lower NDCG@$k = 1$ of 0.972.

On the 571{,}374 post-deployment certificate set, ranking degrades at small values of $k$ and recovers with depth; rising from NDCG@$k = 1$ to aggregate NDCG of 0.9890 (Decision Tree) and 0.9891 (Extra Trees), and remaining above 0.959 for all $k > 10^{4}$. The behavior at $k = 1$ reflects the linear gain rather than a ranking failure: the highest true risk score in the post-deployment set is 4{,}524, and all three models rank another certificate with a true score of 3{,}207 which gives $3{,}207 / 4{,}524 = 0.709$. Both certificates exceed the critical threshold of 1000 by more than a factor of three, so both are escalated immediately and the models' triaging capabilities remain robust.

\subsubsection{Post-deployment Validation on Newer Certificates}\label{sec:postdeploy}

The evaluations above establish that the triage layer ranks well on data drawn from its own collection window. A triage layer is meant to run unattended between retraining cycles, so it also has to hold up as the inventory turns over. We therefore evaluated the tuned models on the 571{,}374 certificates collected August 5--9, 2026, roughly thirteen months after the most recent collection in the modeling corpus and never used in feature engineering, seed selection, or hyperparameter tuning. Each model used the feature configuration selected for it in Section~\ref{sec2}, and no model was refit, recalibrated, or re-tuned. We call this the post-deployment set because it reproduces the position of an operator who trained a triage model once and utilizes the same model over an extended period of time. Both evaluation sets were scored by the same deterministic rubric described in Section~\ref{sec:method}, so the comparison that follows holds the label function fixed and isolates changes in the certificate population rather than changes in how risk is measured. Collection and scoring were separate steps: the certificates were retrieved during August 5--9, 2026 and the deterministic checker was run over them afterwards, so dates reported for the collection and for the scoring run differ.

The right-hand columns of Table~\ref{tab21} report the outcome, and the approach holds. Every model retains an $R^2$ of at least 0.814, seven of eleven retain $R^2 \geq 0.915$, and eight of eleven keep MAE in the single digits. For those eight, a mean absolute error between 4.5 and 7.4 risk-score units sits far below the 1000-point threshold at which we define a Critical defect. These are regression metrics; Section~\ref{sec:ranking} reports the corresponding ranking metrics on the same set, where aggregate NDCG falls by less than one percentage point but shallow-depth ranking degrades materially. The remainder of this subsection analyses the predictive degradation and its causes.

The accuracy degradation we observe is uneven. MAE rises by a factor of 1.86 for the best-preserved model and 4.77 for the worst. The spread matters more than the average because it reorders the test-set ranking. Relative to Random Forest, Extra Trees is 3.6\% more accurate on the test set (2.263 against 2.348) and 49.5\% less accurate on the post-deployment set (6.720 against 4.495), falling from first place to sixth. CatBoost, effectively tied for best on the test set at 2.357, degrades to 11.184, the second worst result in the study. A practitioner selecting a triage model on test-set performance alone would have chosen badly.

\begin{figure}[h]
  \centering
  \includegraphics[width=0.78\linewidth]{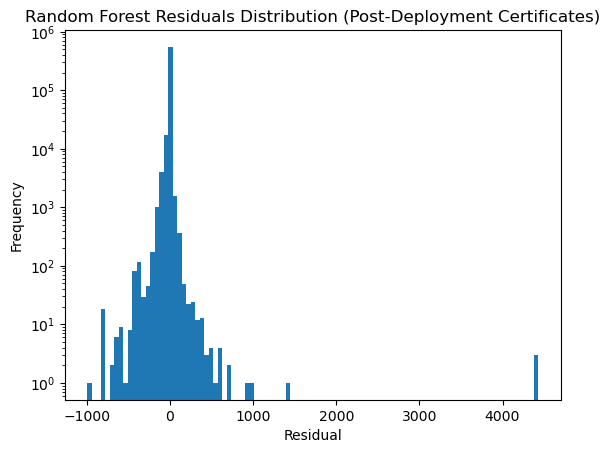}\\[-1mm]
  \includegraphics[width=0.78\linewidth]{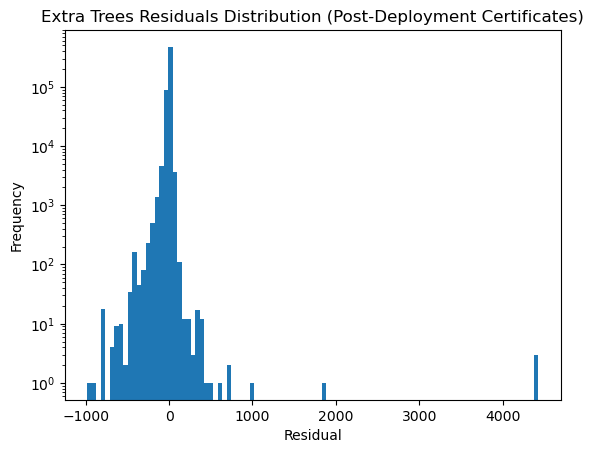}\\[-1mm]
  \includegraphics[width=0.78\linewidth]{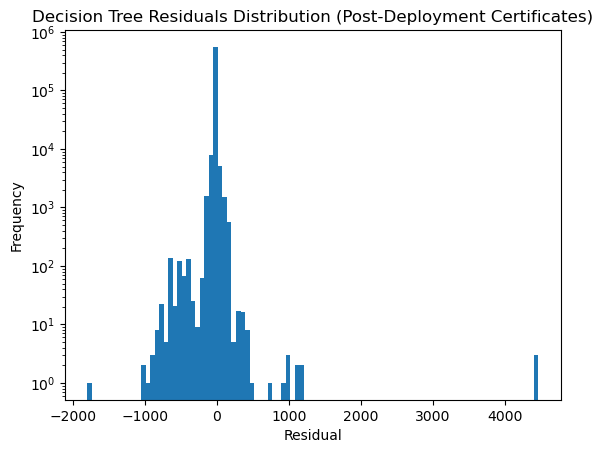}
  \caption{Post-deployment residuals $r = y_t - \hat{y}$ for Random Forest (top),
  Extra Trees (middle), and Decision Tree (bottom) on the 571{,}374 certificates
  collected August 5--9, 2026, with log-scaled frequency. The cluster near
  $r \approx +4400$ appears in all three panels: the highest-risk certificates in
  the set are under-predicted by every model.}
  \label{fig:postdeploy_residuals}
\end{figure}

Fig.~\ref{fig:postdeploy_residuals} shows post-deployment residuals for Random Forest, the strongest model on this set, alongside Extra Trees and Decision Tree, the two models whose test-set residuals were examined in Section~\ref{sec:residuals}. All three panels share the shape reported there: a dominant spike at zero, a body of errors within $\pm$1000, and sparse tails. They differ in the shoulders. Random Forest and Extra Trees keep nearly all non-zero mass within $\pm$600 and place only a handful of certificates beyond $\pm$1000. The Decision Tree's distribution is visibly heavier on the over-prediction side, with sustained mass across the $-1000$ to $-250$ band and one certificate near $-1800$: the tree assigns roughly a 1000-point defect penalty to certificates that do not carry one, which in a triage workflow is a false positive costing an analyst a deterministic re-check.

All three models share one failure that deserves emphasis, and it qualifies the critical-tier recall reported in Section~\ref{sec:tier_eval}. A small cluster of certificates, three in each panel, sits at a residual near $+4400$, an under-prediction exceeding four times the 1000-point critical threshold. These are among the highest-risk certificates in the post-deployment set, and every model misses them by more than the full width of the Critical band, so they would be ranked as unremarkable rather than escalated. critical-tier recall of 98.90\% on the held-out test set therefore should not be read as a guarantee that holds as the inventory turns over. Three certificates out of 571{,}374 is a small absolute count, but the severity argues for pairing the ranked queue with unconditional deterministic checks on the few defect classes whose consequences do not tolerate a probabilistic miss, such as RSA moduli factorable by known techniques.

Severity-tier classification, recomputed on the post-deployment set under the same tier boundaries used in Section~\ref{sec:tier_eval}, degrades far less than point prediction. Table~\ref{tab:tier_postdeploy} reports per-tier metrics for the three models examined here. Overall tier accuracy falls by at most a quarter of a percentage point, from 99.76\% on the held-out test set to 99.52\% (Decision Tree), 99.69\% (Random Forest), and 99.70\% (Extra Trees). Critical-tier recall, the metric a triage workflow depends on, holds at 98.06\% for Decision Tree against 98.90\% on the test set, and at 97.03\% for Extra Trees against 98.79\%. Random Forest is the exception: its critical recall falls to 90.83\%, missing 71 of the 774 critical certificates where Decision Tree misses 15. Tier placement is therefore more durable than either regression error or ranking at shallow depths, which follows from the mechanism: tier assignment depends only on which side of a threshold a prediction falls, so it tolerates ordering errors that a rank-sensitive metric penalizes.

\begin{table}[h]
\centering
\begin{threeparttable}
\caption{Post-Deployment Severity-Tier Classification Metrics}
\label{tab:tier_postdeploy}
\footnotesize
\setlength{\tabcolsep}{4pt}
\begin{tabular}{|l|l|c|c|c|}
\hline
\textbf{Model} & \textbf{Tier} & \textbf{Precision} & \textbf{Recall} & \textbf{F1} \\
\hline
Decision Tree & Low       & 99.64\% & 99.85\% & 99.75\% \\
              & High Sev. & 97.60\% & 94.43\% & 95.99\% \\
              & Critical  & 95.71\% & \textbf{98.06\%} & 96.87\% \\
\hline
Random Forest & Low       & 99.85\% & 99.83\% & 99.84\% \\
              & High Sev. & 97.23\% & 97.66\% & 97.45\% \\
              & Critical  & 97.77\% & 90.83\% & 94.17\% \\
\hline
Extra Trees   & Low       & 99.86\% & 99.83\% & 99.84\% \\
              & High Sev. & 97.33\% & 97.80\% & 97.57\% \\
              & Critical  & 97.66\% & 97.03\% & 97.34\% \\
\hline
\end{tabular}
\begin{tablenotes}[flushleft]\footnotesize
\item Computed on the 571{,}374 post-deployment certificates using the tier boundaries of Section~\ref{sec:tier_eval}: Low ($<100$), High Severity ($100$--$999$), Critical ($\geq 1000$). Models were applied as deployed, with no retraining.
\item Support is identical across models because tiers derive from ground-truth scores: Low 535{,}576, High Severity 35{,}024, Critical 774.
\item Overall tier accuracy: Decision Tree 99.52\%, Random Forest 99.69\%, Extra Trees 99.70\%.
\item Critical-tier support is 0.135\% of this set against 0.94\% of the held-out test set, so the critical row rests on a thinner sample: each additional miss moves recall by 0.13 points.
\end{tablenotes}
\end{threeparttable}
\end{table}

These results change which model we recommend as the triage layer, relative to the test-set reading in Section~\ref{sec:tradeoffs}. Extra Trees, the test-set accuracy leader, is dominated after the shift: Random Forest is 33\% more accurate (MAE 4.495 against 6.720) and 2.23$\times$ faster, so we no longer recommend Extra Trees. Between Random Forest and Decision Tree, Random Forest holds the lower aggregate error, at MAE 4.495 and $R^2$ of 0.941 against 5.204 and 0.915, an advantage of 0.71 risk-score units on a scale where critical defects begin at 1000. Decision Tree recovers 98.06\% of critical-tier certificates against Random Forest's 90.83\%, missing 15 of 774 where Random Forest misses 71, and it does so 50$\times$ faster, at 270 nanoseconds against 13{,}640. We therefore recommend Decision Tree as the triage layer.

\subsection{Feature Importance Analysis}
We used the Extra Trees and Decision Tree models to evaluate which certificate attributes most strongly influence predicted risk scores. The top five Extra Trees features, listed in Table~\ref{tab7},
accounted for 66\% of the total model importance. Certificate validity period and the presence of a TLS Server Authentication EKU (a binary indicator for OID 1.3.6.1.5.5.7.3.1) together contributed
40.1\% of the predictive power, confirming that short validity and correct key usage constraints are indicators of well-managed certificates. Negative serial number presence, no issuer country, and self-signed status were also strong predictors of higher risk.
The TLS Server Authentication EKU's high importance reflects our dataset's composition (85.6\% publicly trusted, predominantly web-facing certificates). In enterprise environments with diverse certificate use cases (e.g., client authentication, code signing, email security), the relative importance of specific EKU types may differ.
Complete feature importance rankings for both models are provided in Appendix~\ref{app:feature_importance} (Tables~\ref{tab:dt_fi} and~\ref{tab:et_fi}). Notably, the Decision Tree concentrates 64.6\% of its importance on validity period alone, while Extra Trees distributes importance more evenly across the top features.

\begin{table}[h]
  \caption{Important Features}
  \centering
  \label{tab7}
  \begin{tabular}{|l|r|}
    \hline
    \textbf{Feature}               & \textbf{Importance (\%)} \\
    \hline
    Validity period in days       & 22.22                   \\
     \hline
    EKU: TLS Server Auth             &  17.92                   \\
    \hline
    Has neg serial number       &  10.21                \\
    \hline
    Issuer country: N/A & 7.92 \\
    \hline
    Is self-signed       &  7.57                 \\
    \hline
  \end{tabular}
\end{table}

\begin{table}[h]
  \caption{Unimportant Features}
  \centering
  \label{tab8}
  \begin{tabular}{|l|r|}
    \hline
    \textbf{Feature}               & \textbf{Importance (\%)} \\
    \hline
    Has registered id      &  0.0005                 \\
    \hline
    Has pseudonym       & 0                   \\
     \hline
    Has unknown                    &  0                  \\
    \hline
    Has edi party name       &  0                 \\
    \hline
    Has x400 address      &  0                \\
    \hline
    % Has data encipherment & 0.39 \\
    % \hline
    % Issuer has dn qualifier & 0.37 \\
    % \hline
    % Has e & 0.23 \\
    % \hline
    % Has l & 0.19 \\
    % \hline
    % Has non-repudiation usage & 0.19 \\
    % \hline
    % Has c & 0.15 \\
    % \hline
    % Has serial number & 0.15 \\
    % \hline
    % Has o & 0.13 \\
    % \hline
    % Has upn & 0.13 \\
    % \hline
    % Has cn & 0.12 \\
    % \hline
    % Issuer has ou & 0.10 \\
    % \hline
    % Issuer has dc & 0.09 \\
    % \hline
    % Issuer has cn & 0.08 \\
    % \hline
    % Key type & 0.07 \\
    % \hline
    % Has crl signing & 0.07 \\
    % \hline
    % Has st & 0.06 \\
    % \hline
    % Issuer has st & 0.05 \\
    % \hline
    % Has title & 0.04 \\
    % \hline
    % Issuer has title & 0.03 \\
    % \hline
    % Has uri & 0.03 \\
    % \hline
    % Has ds replication & 0.03 \\
    % \hline
    % Has dc & 0.02 \\
    % \hline
    % Has email address & 0.02 \\
    % \hline
    % Has entrust & 0.01 \\
    % \hline
    % Has digicert & 0.009 \\
    % \hline
    % Has globalsign & 0.007 \\
    % \hline
    % Has other name & 0.006 \\
    % \hline
    % Has street & 0.005 \\
    % \hline
    % Issuer has uid & 0.002 \\
    % \hline
    % Has registered id & 0.0005 \\
    % \hline
    % Has sn & 0 \\
    % \hline
    % Has g & 0 \\
    % \hline
    % Has uid & 0 \\
    % \hline
    % Has initials & 0 \\
    % \hline
    % Has generation qualifier & 0 \\
    % \hline
    % Has pseudonym & 0 \\
    % \hline
    % Issuer has sn & 0 \\
    % \hline
    % Issuer has street & 0 \\
    % \hline
    % Issuer has g & 0 \\
    % \hline
    % Issuer has initials & 0 \\
    % \hline
    % Issuer has generation qualifier & 0 \\
    % \hline
    % Issuer has pseudonym & 0 \\
    % \hline
    % Has x400 address & 0 \\
    % \hline
    % Has edi party name & 0 \\
    % \hline
    % has unknown & 0 \\
    % \hline
  \end{tabular}
\end{table}

% \subsection{Ablation Study}

\section{Conclusion}

We presented \textit{X-amine509}, an ML-driven approach that predicts enterprise X.509 certificate risk from certificate-visible attributes alone.
Using a corpus of 1{,}027{,}714 certificates across commercial and government domains, we showed that classical tree-based models, strengthened by
targeted feature engineering and tuning, can approximate in-depth deterministic risk analysis at high fidelity and low latency.
On a held-out set of 201{,}976 certificates, our best models achieved MAE of 2.26--2.50 and $R^2$ of 0.993, while our Decision Tree model
offered exceptional performance-speed trade-offs (MAE: 2.395, $R^2$: 0.986) suitable for real-time triage.
When predictions are mapped to severity tiers, the Decision Tree correctly classifies 99.76\% of certificates with 98.90\% recall on critical-tier defects, ensuring that virtually all certificates with critical vulnerabilities are surfaced for deterministic review. Aggregate NDCG of 0.998 confirms that the predicted ordering preserves ground-truth risk priority. Tier placement proved the most durable of these properties: applied without retraining to certificates collected thirteen months later, the Decision Tree retained 98.06\% critical-tier recall while its regression error rose by a factor of 2.17.

We make practical security analysis tractable at scale (Section~1). The Decision Tree model reduces all of this to a handful of ``compare'' and ``jump'' instructions on a single machine. Rather than spending compute resources confirming that the majority of compliant certificates have no issues, ML triage directs deterministic analysis where it matters most.

Beyond speed and accuracy, the results are operationally useful. The models rank certificates by practical risk, highlighting hygiene and policy gaps surfaced
in our EDA (e.g., validity periods, missing key usages, profile enforcement signals). This lets PKI teams prioritize remediation without first running a full
suite of checks across massive inventories. In practice, the models can (i) pre-screen large populations, (ii) route only higher-risk items to expensive
deterministic analysis, and (iii) provide continuous risk monitoring as inventories change. Crucially, the feature importance analysis (Section~4.7) provides coarse triage-level explanations for why a certificate was flagged, while all detailed, actionable reports (specific rule violations and remediation guidance) come from the downstream deterministic checker.

This work has several limitations that should inform deployment decisions.

\textit{Rubric coverage.}
Like any standards-based assessment, our approach is bounded by the defects enumerated in the rubric. Novel attack vectors not yet reflected in PKI standards or security best practices will not be captured until the rubric is updated. However, this limitation applies equally to deterministic tools, which also check only predefined rules. The ML model can be retrained as the rubric evolves to incorporate newly discovered threats.

\textit{Dataset composition and generalizability.}
Our dataset is predominantly publicly trusted, web-facing certificates (85.6\%). Learned feature importances, particularly the prominence of the TLS Server Authentication EKU, may not generalize directly to enterprise inventories with substantial client authentication, code signing, or internal PKI populations. Furthermore, CA/Browser Forum Baseline Requirements, which inform many of our checks, do not govern private enterprise CAs; scores assigned to privately rooted certificates (6.2\% of our dataset) may overestimate risk where enterprises have legitimate policy deviations such as longer validity periods for internal services. Our models predict a composite risk score based on a specific standards-based rubric (Section~3), not individual defect causality or an objective measure of security risk. Features capture what certificates expose, not deployment context such as network topology, key storage, or revocation infrastructure. The contribution is the efficiency of this approximation; organizations should adapt the rubric to their own policies before deployment.  

\textit{False negatives.}
Because the model learns from the distribution of defects in training data, rare but critical defect patterns that are underrepresented in the training set may receive lower predicted risk scores than warranted. The two-stage architecture mitigates this: certificates that the ML model ranks as low-risk but that contain rare critical defects will still be identified when deterministic analysis reaches them in the queue. However, in a deployment where only the top-$k$ ranked certificates receive deterministic review, such certificates could be missed.

These limitations do not undercut the core finding: a compact, interpretable feature set supports fast, accurate prediction of enterprise-relevant certificate risk at scale.

Future directions:
\begin{itemize}
  \item Train defect-specific heads to predict both overall risk and key underlying factors for explainability.
  \item Integrate online learning to adapt to policy changes and new lint rules; add drift detection.
  \item Measure the system's performance in real-world deployments, including latency, accuracy, and explainability.
  \item Refine the set of deterministic checks used in the model.
  \item Continue to explore implications in light of crucial technology developments like autonomous AI agents and practical quantum computers.
  \item Explore incident-based training: an alternative approach would train ML models on incident-derived labels, i.e., certificates whose weaknesses were actually exploited in real-world attacks. This would shift the learning objective from ``how far does this certificate deviate from standards?'' to ``how likely is this certificate to be involved in a security incident?'' We consider this a complementary research direction that requires an incident dataset not currently available in the public domain. Our standards-based approach provides a practical alternative grounded in established PKI security requirements.
\end{itemize}

In sum, \textit{X-amine509} turns certificate-visible attributes into a fast, defensible proxy for full deterministic assessment. It provides a practical path to cut compute costs, surface the riskiest certificates first, and raise the baseline of PKI hygiene in large enterprises.
\newpage
\bibliography{references}

\appendices
\section{Complete Feature Importances}\label{app:feature_importance}

Tables~\ref{tab:dt_fi} and~\ref{tab:et_fi} report the full feature importance rankings for the Decision Tree and Extra Trees models, respectively.
All 81 (Decision Tree) and 88 (Extra Trees) non-zero features are shown.
Of the 102 input features, 21 (Decision Tree) and 14 (Extra Trees) had zero importance and are listed below each table.

\newsavebox{\dtbox}
\begin{table*}[ht]
\caption{Decision Tree Feature Importances (All 81 Non-Zero Features)}
\label{tab:dt_fi}
\centering
\scriptsize
\setlength{\tabcolsep}{3pt}
\savebox{\dtbox}{%
\begin{tabular}{l r @{\hspace{12pt}} l r}
\hline
\textbf{Feature} & \textbf{Imp.\ (\%)} & \textbf{Feature} & \textbf{Imp.\ (\%)} \\
\hline
Validity period (days)    & 64.635 & Issuer has serial no.   & 0.018 \\
Has key cert.\ signing    & 8.022  & Has URI                 & 0.016 \\
Negative serial number    & 6.092  & Has DS replication       & 0.014 \\
Is self-signed            & 3.793  & Issuer ctry: CA          & 0.012 \\
Key size (bits)           & 3.135  & Issuer has ST            & 0.011 \\
Serial number length      & 3.062  & SAN count: UPN           & 0.010 \\
Signing algorithm         & 1.869  & Has ST                   & 0.009 \\
SAN count: DNS            & 1.587  & Issuer has L             & 0.008 \\
Has E                     & 1.531  & Issuer has title         & 0.008 \\
Issuer has O              & 1.095  & Has directory name       & 0.007 \\
EKU 3                     & 0.809  & Issuer ctry: US          & 0.006 \\
Certificate version       & 0.673  & Has L                    & 0.006 \\
Number of SANs            & 0.511  & Has UPN                  & 0.005 \\
Issuer has C              & 0.410  & Issuer has DN qualifier  & 0.005 \\
Has data encipherment     & 0.388  & Key type                 & 0.005 \\
EKU 0 (serverAuth)        & 0.270  & SAN count: email         & 0.004 \\
Has digital signature     & 0.265  & SAN count: URI           & 0.004 \\
SAN count: DirName        & 0.251  & Has email address        & 0.003 \\
Has key agreement         & 0.219  & Issuer ctry: GB          & 0.003 \\
SAN count: IP             & 0.128  & Issuer ctry: other       & 0.002 \\
Has C                     & 0.127  & Is Entrust               & 0.002 \\
EKU 1                     & 0.112  & Issuer ctry: UK          & 0.002 \\
Has serial number         & 0.104  & Is DigiCert              & 0.001 \\
Has CRL signing           & 0.093  & EKU 2                    & $<$0.001 \\
Has DNS name              & 0.091  & Issuer ctry: AU          & $<$0.001 \\
SAN count: DS repl.       & 0.051  & Issuer ctry: BE          & $<$0.001 \\
Has no key usage          & 0.049  & Issuer ctry: JP          & $<$0.001 \\
Issuer has DC             & 0.048  & Issuer ctry: IN          & $<$0.001 \\
Has O                     & 0.047  & Issuer has UID           & $<$0.001 \\
Has OU                    & 0.046  & Issuer has street        & $<$0.001 \\
SAN count: otherName      & 0.043  & Is Hydrant               & $<$0.001 \\
Issuer has E              & 0.042  & Issuer ctry: FR          & $<$0.001 \\
Has IP address            & 0.036  & Has UID                  & $<$0.001 \\
Has key encipherment      & 0.034  & Issuer ctry: N/A         & $<$0.001 \\
Has CN                    & 0.033  & Issuer ctry: NO          & $<$0.001 \\
Issuer has OU             & 0.027  & Is GlobalSign            & $<$0.001 \\
Has title                 & 0.025  & Issuer ctry: CN          & $<$0.001 \\
Has DC                    & 0.023  & EKU 4                    & $<$0.001 \\
Issuer ctry: empty        & 0.023  & Has street               & $<$0.001 \\
Issuer has CN             & 0.020  & Issuer ctry: AT          & $<$0.001 \\
Has non-repudiation       & 0.018  &                          &        \\
\hline
\end{tabular}}%
\usebox{\dtbox}

\vspace{2pt}
\begin{minipage}{\wd\dtbox}
\raggedright\scriptsize
\textit{21 zero-importance features (omitted):}
has\_sn, has\_g, has\_initials, has\_generation\_qualifier, has\_pseudonym, has\_dn\_qualifier,
has\_registered\_id, has\_other\_name, has\_x400\_address, has\_edi\_party\_name, has\_unknown,
is\_encipher\_only, is\_decipher\_only, is\_from\_NL, issuer\_has\_sn, issuer\_has\_g,
issuer\_has\_initials, issuer\_has\_generation\_qualifier, issuer\_has\_pseudonym,
san\_count\_edi\_party, san\_count\_registered\_id.
\end{minipage}
\end{table*}

\newsavebox{\etbox}
\begin{table*}[ht]
\caption{Extra Trees Feature Importances (All 88 Non-Zero Features)}
\label{tab:et_fi}
\centering
\scriptsize
\setlength{\tabcolsep}{3pt}
\savebox{\etbox}{%
\begin{tabular}{l r @{\hspace{12pt}} l r}
\hline
\textbf{Feature} & \textbf{Imp.\ (\%)} & \textbf{Feature} & \textbf{Imp.\ (\%)} \\
\hline
Validity period (days)    & 22.223 & Has CRL signing          & 0.055 \\
EKU 0 (serverAuth)        & 17.925 & Issuer ctry: NL          & 0.055 \\
Negative serial number    & 10.209 & Key type                 & 0.051 \\
Issuer ctry: N/A          & 7.917  & Issuer ctry: IN          & 0.050 \\
Is self-signed            & 7.578  & Has CN                   & 0.041 \\
Has IP address            & 5.117  & SAN count: UPN           & 0.040 \\
Has key cert.\ signing    & 3.687  & Issuer has title         & 0.039 \\
Serial number length      & 3.042  & Issuer has ST            & 0.033 \\
Certificate version       & 2.700  & Has title                & 0.032 \\
Issuer ctry: empty        & 2.462  & Has UPN                  & 0.032 \\
Key size (bits)           & 2.277  & Has ST                   & 0.031 \\
Has non-repudiation       & 1.992  & SAN count: DS repl.      & 0.028 \\
Has OU                    & 1.744  & Has DS replication       & 0.027 \\
Issuer has DN qualifier   & 1.621  & Issuer has CN            & 0.021 \\
Has key agreement         & 1.530  & Has DC                   & 0.021 \\
Signing algorithm         & 1.030  & Has URI                  & 0.014 \\
Has DNS name              & 0.512  & SAN count: URI           & 0.011 \\
Has digital signature     & 0.483  & Issuer ctry: FR          & 0.010 \\
Has no key usage          & 0.435  & Has street               & 0.009 \\
Issuer ctry: UK           & 0.421  & Issuer ctry: CN          & 0.008 \\
Has E                     & 0.397  & Is GlobalSign            & 0.008 \\
Has data encipherment     & 0.329  & Issuer ctry: other       & 0.006 \\
Has serial number         & 0.328  & SAN count: email         & 0.005 \\
Issuer has serial no.     & 0.317  & Has email address        & 0.005 \\
Has key encipherment      & 0.302  & EKU 2                    & 0.005 \\
Issuer has E              & 0.296  & Is DigiCert              & 0.003 \\
Number of SANs            & 0.280  & Is Entrust               & 0.002 \\
Issuer has L              & 0.274  & Issuer ctry: GB          & 0.002 \\
Issuer has O              & 0.274  & Issuer has UID           & 0.001 \\
SAN count: DNS            & 0.251  & SAN count: otherName     & 0.001 \\
Has DN qualifier          & 0.174  & Has other name           & 0.001 \\
EKU 3                     & 0.144  & Issuer ctry: JP          & $<$0.001 \\
Issuer has OU             & 0.142  & Issuer ctry: AU          & $<$0.001 \\
Issuer ctry: US           & 0.126  & EKU 4                    & $<$0.001 \\
Has C                     & 0.121  & Has UID                  & $<$0.001 \\
EKU 1                     & 0.108  & Issuer has street        & $<$0.001 \\
SAN count: IP             & 0.104  & Issuer ctry: BE          & $<$0.001 \\
Has L                     & 0.090  & Has registered ID        & $<$0.001 \\
Issuer has C              & 0.080  & SAN count: registered ID & $<$0.001 \\
SAN count: DirName        & 0.068  & Issuer ctry: NO          & $<$0.001 \\
Has directory name        & 0.064  & Is Hydrant               & $<$0.001 \\
Has O                     & 0.062  & Is encipher only         & $<$0.001 \\
Issuer has DC             & 0.058  & Issuer ctry: AT          & $<$0.001 \\
Issuer ctry: CA           & 0.056  & Is decipher only         & $<$0.001 \\
\hline
\end{tabular}}%
\usebox{\etbox}

\vspace{2pt}
\begin{minipage}{\wd\etbox}
\raggedright\scriptsize
\textit{14 zero-importance features (omitted):}
has\_sn, has\_g, has\_initials, has\_generation\_qualifier, has\_pseudonym, has\_unknown,
has\_x400\_address, has\_edi\_party\_name, issuer\_has\_sn, issuer\_has\_g, issuer\_has\_initials,
issuer\_has\_generation\_qualifier, issuer\_has\_pseudonym, san\_count\_edi\_party.
\end{minipage}
\end{table*}

\bibliographystyle{IEEEtran}
\end{document}